\documentclass[useAMS,usenatbib]{mn2e} 
\usepackage{aas_macros}
\usepackage{graphics}
\usepackage{epsfig}  
\usepackage{natbib} 
\usepackage{float}
\usepackage{amsmath}
\usepackage{times}
\usepackage[varg]{txfonts}
\newcommand{\amiga}{\texttt{AMIGA}}

\newcommand{\mlapm}{\texttt{MLAPM}}
\newcommand{\hMpc}{{\ifmmode{h^{-1}{\rm Mpc}}\else{$h^{-1}$Mpc}\fi}}
\newcommand{\hkpc}{{\ifmmode{h^{-1}{\rm kpc}}\else{$h^{-1}$kpc}\fi}}
\newcommand{\hMsun}{{\ifmmode{h^{-1}{\rm {M_{\odot}}}}\else{$h^{-1}{\rm{M_{\odot}}}$}\fi}}

\def\lesssim{\mathrel{\hbox{\rlap{\hbox{\lower4pt\hbox{$\sim$}}}\hbox{$<$}}}}
\def\gtrsim{\mathrel{\hbox{\rlap{\hbox{\lower4pt\hbox{$\sim$}}}\hbox{$>$}}}}

\title[Cosmological Magnetohydrodynamics with \texttt{AMIGA}]
      {Investigating the influence of magnetic fields upon structure formation with AMIGA -- a C code for cosmological magnetohydrodynamics}
\author[T. Doumler and A. Knebe] 
{Timur Doumler$^{1,2}$\thanks{E-mail: tdoumler@aip.de} and Alexander Knebe$^{3}$   \\
  $^1$Universit\'e Lyon 1, CNRS/IN2P3/INSU, Institut de Physique Nucl\'eaire, 69622 Villeurbanne, Lyon, France\\
  $^2$Astrophysikalisches Institut Potsdam, An der Sternwarte 16, 14482 Potsdam, Germany\\
  $^3$Departamento de F\'\i sica Te\'orica, Modulo C-15, Facultad de Ciencias, Universidad Aut\'onoma de Madrid, 28049 Cantoblanco, Madrid, Spain
  }

\begin{document}

\date{accepted by MNRAS 2009 December 1.}

\pagerange{\pageref{firstpage}--\pageref{lastpage}} \pubyear{2008}

\maketitle

\label{firstpage}

\begin{abstract}
Despite greatly improved observational methods, the presence of magnetic fields at cosmological scales and their role in the process of large-scale structure formation still remains unclear. In this paper we want to address the question how the presence of a hypothetical primordial magnetic field on large scales influences the cosmic structure formation in numerical simulations. As a tool for carrying out such simulations, we present our new numerical code \amiga. It combines an $N$-body code with an Eulerian grid-based solver for the full set of MHD equations in order to conduct simulations of dark matter, baryons and magnetic fields in a self-consistent way in a fully cosmological setting. Our numerical scheme includes effective methodes to ensure proper capturing of shocks and highly supersonic flows and a divergence-free magnetic field. The high accuracy of the code is demonstrated by a number of numerical tests. We then present a series of cosmological MHD simulations and confirm that, in order to have a significant effect on the distribution of matter on large scales, the primordial magnetic field strength would have to be significantly higher than the current observational and theoretical constraints.
\end{abstract}


\section{Introduction}
\label{sec:introduction}

Today, we arrived at an era where cosmology has finally reached the stage of a precision science. The cosmological parameters have been determined to a typical precision of very few percents, resulting in the standard $\Lambda$CDM model of cosmology \citep{Komatsu09}, and in this context the process of the cosmic structure formation can be studied in great detail. The non-linear nature of the gravitational dynamics and gas physics make the problem of structure formation virtually intractable analytically, and therefore the field relies on numerical simulations, which have been the driving force behind much of the theoretical progress. While the first codes were just able to follow the evolution of dark matter with the $N$-body method \citep[e.g.][]{Klypin83,Efstathiou85, Davis85,Barnes86,Villumsen89,Couchman91,Suisalu95,Kravtsov97,Knebe01}, the tremendous increase in computational power in the last years made it possible to include more and more different components into the simulations \citep[e.g.][]{Couchman95,Teyssier02,OShea04,Springel05,Li08,Xu08,Collins09}. The inclusion of baryon physics, star formation, AGN feedback, radiative transfer and magnetic fields in modern cosmological codes opens the way to study many different aspects of the structure formation process.

Magnetic fields play an important role in astrophysical phenomena on many different scales. Since most of the visible matter in the universe is ionized and magnetic fields are found on every scale where they can be observed, it is natural that they could also play a cosmological role. Unfortunately, magnetic fields on scales larger than individual galaxies are much more difficult to observe -- any measurement of magnetic fields must rely on the presence of radiation and a magnetized medium. So far, the largest-scale observable magnetic fields are inside the atmospheres of galaxy clusters \citep{Carilli02,Govoni04}, reaching strengths of the order of $\mu$G in the core regions. Detection methods include studies of radio synchrotron and inverse Compton X-ray emission from clusters \citep{Harris79,Rephaeli87} and surveys of  Faraday rotation measures of polarized radio sources passing through the cluster atmosphere \citep{Clarke01}.  However, it could be possible that there are magnetic fields on even larger scales than the largest observable objects. In fact, all the ``empty space'' in the universe could be magnetized \citep{Kronberg99}. A truly cosmological magnetic field would not be associated with collapsing or gravitationally bound structures, and would be coherent on scales greater than the largest known structures ($\sim 100$ Mpc) or even the Hubble radius, permeating the whole universe. Although a new generation of highly sensitive radio telescopes like LOFAR and SKA is underway, the detection of such a cosmological field will probably stay out of reach for the next decades. Currently, it is only possible to estimate an upper limit for the strength of such a large-scale field \citep{Vallee90,Kronberg94}, which should not be higher than $\sim$ 1 nG. In order to learn more about the nature and effects of magnetic fields, we have to rely on theoretical models and numerical simulations. Even if we may not be able to directly prove the existence of a large-scale magnetic field, the subject has important cosmological implications that must be considered. A large-scale magnetic field can have a significant impact on the dynamics of cosmic baryon flows, the thermal and ionization history of the universe, and the onset of structure formation \citep{Sethi08}.

Different theories exist on the origin of a cosmological magnetic field. One class of theories suggests that the creation of a universal, ``primordial'' magnetic field happened already during a very early stage of the evolution of the universe. Unfortunately, at present such theories are highly parameter-dependent and rather inconclusive \citep{Subramanian08}, and they do not yet allow to derive the field strength of such a primordial field; it is currently only possible to estimate some upper limit. Again, numerical simulations seem to be a very good alternative to learn more about this subject.

Cosmological simulations including magnetic fields have already been conducted during the last decade. It has been shown by numerical simulations that indeed a large-scale primordial field of order $\sim$ 1 nG is needed to explain the presence of the observed magnetic fields in galaxy clusters \citep{Dolag99}. There have also been simulations of magnetic fields in filaments \citep{Brueggen05}, cosmological simulations studying cosmic-ray electrons \citep{Miniati01}, and the influence of magnetic pressure on the growth of baryonic structures \citep{Gazzola07}. However, aside from \citet{Dolag99}, these earlier codes only included magnetic fields passively, neglecting any possible back-reaction effects on the baryons; or, as in \citet{Gazzola07}, the magnetic component was modelled simply by adding an additional, isotropic pressure term in the hydrodynamic equations to account for the magnetic pressure.

In order to track magnetic fields, baryons and dark matter simultaneously in a self-consistent way inside a cosmological framework, it is necessary to numerically solve the full set of equations of cosmological magnetohydrodynamics (MHD). Codes capable of this task have started to be developed only very recently. They include grid codes \citep{Fromang06, Li08, Collins09} as well as an SPH code \citep{Dolag09}, none of which were publicly available at the time of writing. In this paper, we present the new cosmological MHD code \amiga, aimed to close this gap.\footnote{\amiga\ can be downloaded from the following web site: \texttt{http://popia.ft.uam.es/AMIGA}.}

The \amiga\ code originally started as a pure $N$-body code to study dark matter structure formation. It is the successor of the \mlapm\ code, a very powerful and memory-efficient AMR code published by \citet{Knebe01}.  Here, we present a new numerical solver for cosmological MHD, now implemented into the \amiga\ code. It greatly improves the possibilities of the code, allowing to model dark matter, baryons and magnetic fields simultaneously in a fully cosmological setting. The code utilizes the transformation to \emph{supercomoving} coordinates, which greatly simplifies the numerical solution of cosmological MHD equations. There are implemented techniques to properly resolve strong shockwaves and supersonic flows in the baryon component, and to ensure the important condition of a divergence-free magnetic field down to machine precision.  We also present a series of test problems, in order to verify the high accuracy of the code.

After a technical description of the underlying principles and numerical methods, we use this new powerful tool to investigate and quantify the influence of a primordial magnetic field on the cosmic structure formation on large scales. Recent numerical efforts in cosmological MHD have concentrated on the modelling of magnetic fields inside individual galaxy clusters, since they are directly observable. For example, \citet{Dubois08} focused specifically on the magnetic field inside one simulated cluster and found a relation between the field strength in the cluster core and cooling processes of the intracluster gas. We want to go a different path and study the influence of a hypothetical universal magnetic field, filling the whole universe, on large-scale structure formation. We currently only have some constraints on the maximum value of such a field, and no working general theory describing its origin or its development until the present time. At this juncture, it seems reasonable to choose a more pragmatic strategy. We want to address an important question for future cosmological simulations: suppose a cosmological primordial field exists -- could it have a dynamically significant influence on the other constituents, dark matter and baryons? Does it need to be considered when performing simulations of the large-scale structure? At what strengths of a primordial field do its effects become relevant, and how do these field strengths compare to the current constraints?  A series of numerical simulations of the large-scale structure formation, conducted with the new cosmological MHD code \amiga\ and including primordial magnetic fields of different strength, is presented here to address these questions.

The outline of this paper is as follows. Section \ref{sec:amiga} is dedicated to our new cosmological MHD code \amiga. We present the supercomoving framework, in which we formulate the equations of ideal MHD (\ref{sec:equations}), the numerical scheme implemented in \amiga\ (\ref{sec:scheme}), and then we carry out different numerical tests to ensure that the code is functioning accurately (\ref{sec:codetesting}). Section \ref{sec:cosmomhd} presents our simulations of structure formation with a primordial large-scale magnetic field. We first use the MHD pancake formation as a toy model to estimate the effect of the fields (\ref{mhdpancake}), and then present our 3D cosmological simulations with magnetic fields and analyse the obtained numerical data (\ref{cosmoMHDsimulations}). Our conclusions are summarized in section \ref{sec:summary}. The derivation of the supercomoving MHD equations is given in the Appendix.

\section{AMIGA}
\label{sec:amiga}

\amiga\ is a cosmological grid code containing the $N$-body solver with adaptive mesh refinement from its predecessor, the \mlapm\ code \citep{Knebe01}, which is used  for the dark matter and gravity equations, and a newly developed MHD solver to track the baryon physics and magnetic fields on a regular grid. 

\subsection{Supercomoving ideal MHD equations}
\label{sec:equations}

Our simulations contain dark matter particles, treated by an $N$-body code, and a baryon component that behaves like an ideal, superconducting plasma, together with a magnetic field. To treat all these components simultaneously in a self-consistent way, the equations of ideal magnetohydrodynamics (MHD) have to be solved together with the dark matter particle equations in a fully cosmological setting. For this, the equations of MHD are usually transformed to the comoving frame, defined by
\begin{align*}
\boldsymbol x= \frac{\boldsymbol r}{a}
\end{align*}

In this frame, in the absence of additional forces, the mass points are at rest and the local density remains constant. However, if applied to the equations of MHD, the transformation renders them into equations with lots of additional factors explicitly depending on the cosmological expansion factor $a(t)$; these equations no longer have the form of hyperbolic conservation laws. Nevertheless, most other cosmological MHD codes use this formulation \citep{Li08,Collins09}. A different transformation to so-called \emph{supercomoving} coordinates has been proposed by \citet{Martel98} to cast the equations into a more convenient form.\footnote{We like to note that codes such as, for instance, \texttt{RAMSES} \citep{Teyssier02} and \texttt{ART} \citep{Kravtsov97, Kravtsov02} also implement supercomoving coordinates. However, neither of the code description papers has yet shown the full set of the corrresponding equations and their derivation as presented here and in the Appendix, respectively.} It is defined in a very similar way, but additionally, the physical time $t$ is replaced by a new function $t_x$ depending on the expansion:
\begin{align}
\label{definition}
\boldsymbol x= \frac{\boldsymbol r}{a}\;; \;\;\;
\textrm{d}t_x=\frac{\textrm{d}t}{a^2}
\end{align}
All time derivatives are now formulated in respect to that new function, and the equations are transformed accordingly. Here, we apply this transformation to the full set of MHD equations. Additionally, the physical quantities therein get substituted by a set of new `supercomoving' quantities:
\begin{align}
\label{transformation}
\rho_x&=\rho a^3 \;\;\;  &T_x&=a^2T\\ \notag
\phi_x&=a^2(\phi+\frac{1}{2}a \ddot a x^2) &S_x&=a^{-(3\gamma-8)}S\\ \notag
p_x&=a^5p &\boldsymbol B_x &= a^{5/2} \boldsymbol B\\ \notag
\varepsilon_x&=a^2\varepsilon &\mathcal H_x&=a\dot a
\end{align}

where $\rho$ is the baryon density, $\phi$ the total gravitational potential, $p$ the thermal baryonic pressure, $\varepsilon$ the thermal baryonic energy, $T$ the temperature, $S$ the modified entropy (definition see section \ref{dualenergy}), $\mathcal H$ the supercomoving Hubble constant and $\boldsymbol B$ the magnetic field strength. With the supercomoving framework defined in that way, the substitution causes most of the $a(t)$ depending terms to cancel out and results in the following equations (the $x$ subscripts are dropped from here on):
\begin{align}
\label{dmx_eq}
&\frac{\textrm{d} \boldsymbol x_{DM}}{\textrm{d}t}=\boldsymbol v_{DM} \displaybreak[0] \\
\label{dmv_eq}
&\frac{\textrm{d} \boldsymbol v_{DM}}{\textrm{d}t}=-\boldsymbol\nabla\phi \displaybreak[0] \\
\label{poisson_eq}
&\Delta\phi=4\pi G(\rho_{tot}-\bar\rho_{tot})\cdot a(t)\displaybreak[0] \\
\label{density_eq}
&\frac{\partial \rho}{\partial t}+\boldsymbol\nabla\cdot(\rho \boldsymbol v)=0\displaybreak[0] \\
\label{momentum_eq}
&\frac{\partial \rho \boldsymbol v}{\partial t}+\boldsymbol\nabla\cdot\left [\rho \boldsymbol v \boldsymbol v + \left(p+\frac{B^2}{2\mu}\right)I-\frac{1}{\mu} \boldsymbol B \boldsymbol B\right]=-\rho\, \boldsymbol\nabla\phi\displaybreak[0] \\
\label{energy_eq}
&\frac{\partial \rho E}{\partial t}+\boldsymbol\nabla\cdot\left[\boldsymbol v \left(\rho E+p+\frac{B^2}{2\mu}\right)-\frac{1}{\mu}\boldsymbol B(\boldsymbol v\cdot\boldsymbol B)\right]\displaybreak[0] \\ \notag
&\;\;=-\rho \boldsymbol v \cdot(\boldsymbol\nabla\phi)+\mathcal H\frac{B^2}{2\mu}\displaybreak[0] \\
\label{induction_eq}
&\frac{\partial \boldsymbol B}{\partial t}+\boldsymbol\nabla\times(-\boldsymbol v \times\boldsymbol B)=\frac{1}{2}\mathcal H\boldsymbol B\displaybreak[0] \\
\label{divB_eq}
&\boldsymbol\nabla\cdot\boldsymbol B=0
\end{align}

The derivation is presented in the appendix. Equations (\ref{dmx_eq}) and (\ref{dmv_eq}) are the equations of motion for the collisionless dark matter (DM) particles, where $\boldsymbol x_{DM}$ is the position and $\boldsymbol v_{DM}$ the velocity, respectively. (\ref{poisson_eq}) is Poisson's equation for the total gravitational potential $\phi$, where $\rho_{tot}$ is the total density of combined gas and dark matter, and $\bar\rho_{tot}$ is the average total density of the simulated box, given by
\begin{align}
\label{average_dens}
\bar\rho_{tot}=\Omega_0\,\rho_{crit}=\Omega_0\frac{3 H_0^2}{8\pi G}
\end{align}
Next there are the supercomoving ideal MHD equations in conservative form, where equation (\ref{density_eq}) is the conservation law for gas density $\rho$, equation (\ref{momentum_eq}) for the gas flow momentum $\rho \boldsymbol v$, and (\ref{energy_eq}) for the total energy density $\rho E$ of the gas. $\boldsymbol B$ is the supercomoving magnetic field, whose evolution is given by the law of induction (\ref{induction_eq}), subject to the divergence-free condition (\ref{divB_eq}). The thermal pressure $p$ is obtained via an ideal equation of state,
\begin{align}
\label{eq_of_state}
p=(\gamma -1)\rho \varepsilon
\end{align}
where the adiabatic index equals $\gamma=5/3$ for a non-relativistic, monoatomic ideal gas, while the internal energy density $\rho\varepsilon$ of the gas follows from the total, kinetic and magnetic energy densities:
\begin{align}
\label{edens}
\rho E=\frac{1}{2}\rho v^2+\frac{B^2}{2}+\rho\varepsilon
\end{align}

Note that formally the equations (\ref{dmx_eq}) to (\ref{induction_eq}) closely resemble their non-comoving counterparts. The only differences are in Poisson's equation for the gravitational potential and the two additonal magnetic Hubble terms at the right side of equations (\ref{energy_eq}) and (\ref{induction_eq}). These are now the only places where cosmology explicitly enters, namely in the form of the supercomoving $a(t)$ function, which has to be determined depending on the adopted cosmological model. These properties of the supercomoving MHD equations make them easier to implement than comoving MHD, while still containing the same physics. In particular they make it very easy to employ numerical schemes originally designed for non-cosmological purposes. 

The code uses the following internal units: The distance unit is the comoving boxsize $B_0$, so that $x,y,z\in[0,1]$ always; the density unit is the average density (\ref{average_dens}), so internally $\delta=\rho-1$; the unit for supercomoving time is the Hubble time $1/H_0$; and the magnetic field unit is defined by setting the magnetic constant to unity: $\mu=1$, so it disappears from all equations. The expansion factor $a(t)$ is evaluated by numerically integrating

\begin{align}
\label{codetimeline}
\frac{\textrm{d}a}{\textrm{d}t}=a^2 \left[ \Omega_\Lambda (a^2-1) + \frac{\Omega_m}{a} - \Omega_m + 1 \right]^{1/2}
\end{align}

internally in the code (this relation results from the Friedmann equation). Note that here, $t$ is the supercomoving time (hence the additional $a^2$) and $H_0\equiv 1$ due to the internal units.

\subsection{Numerical scheme}
\label{sec:scheme}

For an elaborate description of the AMR solver for equations~(\ref{dmx_eq})-(\ref{poisson_eq}) we refer the reader to the \texttt{MLAPM} paper by \citet{Knebe01}. Below, we present the  new solver for the MHD equations~(\ref{density_eq})-(\ref{divB_eq}).

\subsubsection{MHD solver}

The MHD solver of \amiga\ serves to solve the cosmological MHD equations (\ref{density_eq}) -- (\ref{divB_eq}). It consists of a second-order unsplit Godunov-type central scheme and a constrained transport scheme to ensure a divergence-free magnetic field down to machine precision. It is essentially an expanded, cosmological version of the solver used by the \textsc{Nirvana} code \citep{Ziegler04,Ziegler05}, which in turn adopts the KNP solver for hyperbolic conservation laws \citep{Kurganov01}.\footnote{We like to note that the KNP flux is equivalent to the HLL flux formula introduced by \citet{Harten83} as two-speed approximate Riemann solver.} In the following section, we  present the numerical algorithm, including the KNP solver, as implemented in \amiga; for more on the theory behind the scheme, we refer the reader to these articles.

The three hydrodynamic conservation laws (\ref{density_eq}), (\ref{momentum_eq}) and (\ref{energy_eq}) can be written in general vector form:
\begin{align}
\label{general_conserv}
\frac{\partial \boldsymbol{u}}{\partial t}+
\frac{\partial \boldsymbol f^x}{\partial x}+
\frac{\partial \boldsymbol f^y}{\partial y}+
\frac{\partial \boldsymbol f^z}{\partial z}=\boldsymbol S_u
\end{align}
where $\boldsymbol{u}$ is a vector containing the hydrodynamic variables,
\begin{align*}
\boldsymbol{u}=\left(
   \begin{array}{c}
   \rho \\
   \rho v_x \\
   \rho v_y \\
   \rho v_z \\
   \rho E
   \end{array}
\right)
\end{align*}
$\boldsymbol{f}^x, \boldsymbol{f}^y, \boldsymbol{f}^z$ are the flux functions
\begin{align}
\label{def_fluxfunc}
\boldsymbol{f}^x=\left(
   \begin{array}{c}
   \rho v_x\\
   \rho v_x^2+p+B^2/2-B_x^2 \\
   \rho v_x v_y - B_x B_y \\
   \rho v_x v_z - B_x B_z \\
   \rho v_x( E + p + B^2/2)-B_x(\boldsymbol vÊ\cdot \boldsymbol B)
   \end{array}
\right)\\ \notag
\boldsymbol{f}^y=\left(
   \begin{array}{c}
   \rho v_y\\
   \rho v_x v_y-B_x B_y \\
   \rho v_y^2 +p+B^2/2- B_y^2 \\
   \rho v_y v_z - B_y B_z \\
   \rho v_y( E + p + B^2/2)-B_y(\boldsymbol vÊ\cdot \boldsymbol B)
   \end{array}
\right)\\ \notag
\boldsymbol{f}^z=\left(
   \begin{array}{c}
   \rho v_z\\
   \rho v_x v_z-B_x B_z \\
   \rho v_y v_z - B_y B_z \\
   \rho v_z^2+p+B^2/2 - B_z^2 \\
   \rho v_z( E + p + B^2/2)-B_z(\boldsymbol vÊ\cdot \boldsymbol B)
   \end{array}
\right)
\end{align}
and $\boldsymbol{S}_u$ are the source terms
\begin{align}
\label{u_source}
\boldsymbol{S}_u=\left(
   \begin{array}{c}
   0\\
   \rho \,\partial_x \phi \\
   \rho \,\partial_y \phi \\
   \rho \,\partial_z \phi \\
   \rho \boldsymbol v \cdot (\boldsymbol\nabla\phi)+HB^2/2
   \end{array}
\right)
\end{align}

\amiga\ was developed from a particle-mesh code and inherited its grid structure. We use the cells of this grid to locally store discrete values of the MHD quantities $\boldsymbol u, \boldsymbol B$.

\begin{figure}
\begin{center}
\begin{minipage}{0.43\textwidth}
        \epsfig{file=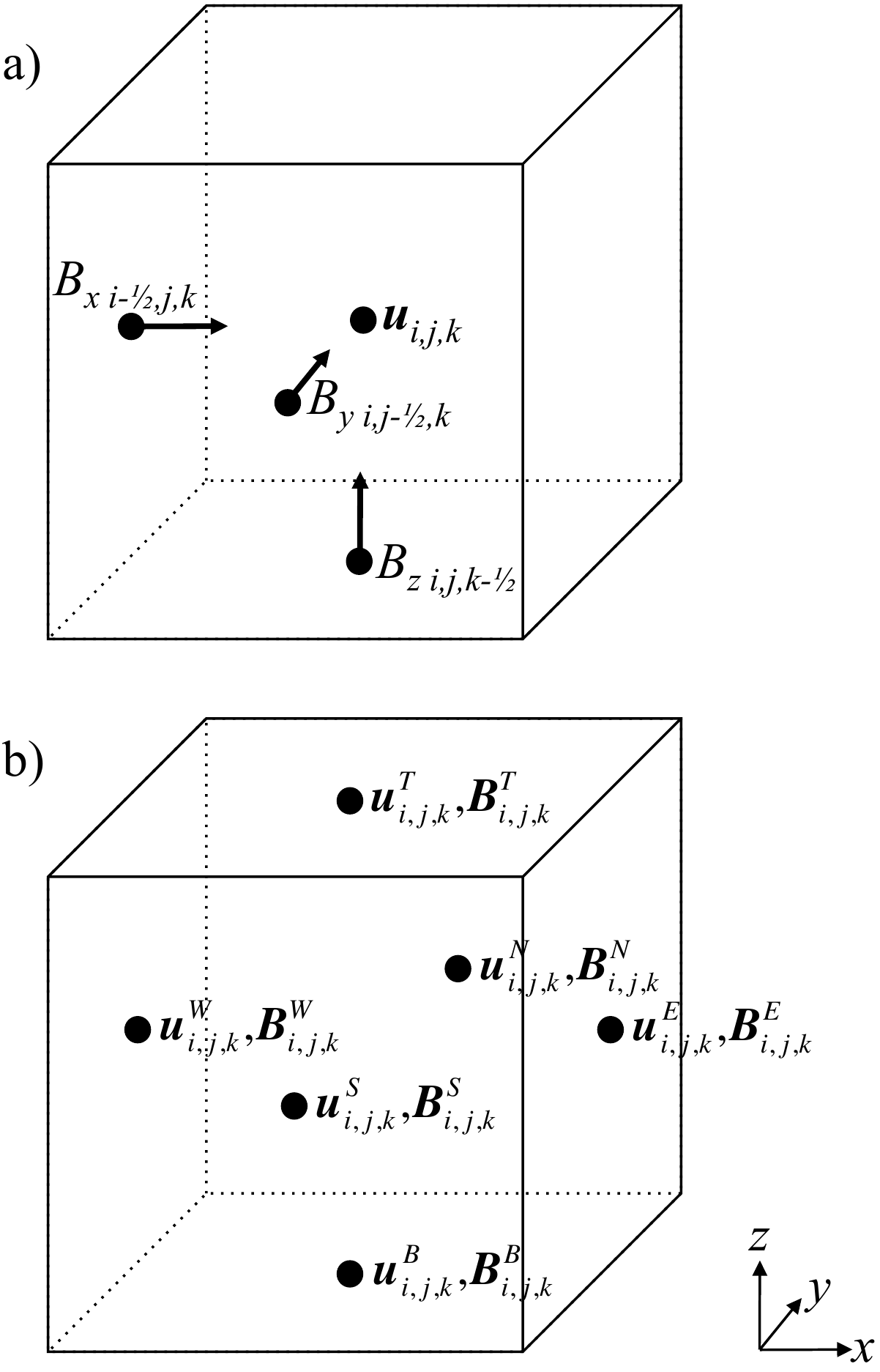, width=1\textwidth, angle=0}
\end{minipage}
\end{center}
\caption{a) The variables stored at every cell $i,j,k$ of the 3D grid. The cell-averaged hydrodynamic quantities $\boldsymbol u_{i,j,k}$ are stored at cell centres $i,j,k$, while the components of $\boldsymbol B$ are stored on different cell faces (``staggered grid''). b) The reconstructed MHD quantities on all six cell faces.
\label{fig:grid}}
\end{figure}

The hydrodynamical quantities $\boldsymbol u$ are stored as cell-averaged values $\boldsymbol u_{i,j,k}$ at the centres of the grid cells $i,j,k$ whereas the magnetic field $\boldsymbol B$ is instead arranged in a ``staggered grid'', i.e. it is stored on the \emph{cell faces} with a staggered collocation of the components $B_x$, $B_y$, $B_z$ (see Figure \ref{fig:grid} a). Thus, every component of the $\boldsymbol B$ field is stored at another interface of the cell.

In three dimensions, the KNP solver requires the reconstruction of all these quantities to all six faces of every cell (see Figure \ref{fig:grid} b). We will denote the six interfaces with the letters $W,E,S,N,T,B$. For example, we obtain a hydrodynamical variable $u$ at the interfaces lying in $x$ direction (E and W interfaces) by
\begin{align}
u_{i,j,k}^E=u_{i,j,k}+\frac{1}{2}(\delta_x u)_{ijk}\\
u_{i,j,k}^W=u_{i,j,k}-\frac{1}{2}(\delta_x u)_{ijk}
\end{align}
where $(\delta_x u)$ is a TVD slope limiter. For cosmological MHD simulations, we use the slope limiter of \citet{vanLeer77}:
\begin{align*}
(\delta_x u)_{ijk}=\frac{2 \max \{(u_{i+1,j,k}-u_{i,j,k})\cdot(u_{i,j,k}-u_{i-1,j,k}),0\}}{u_{i+1,j,k}-u_{i-1,j,k}}
\end{align*}
For the $y$ direction (N and S interfaces) and the $z$ direction (T and B interfaces) the formulae are analogue.

Since there are magnetic terms present in the hydrodynamic flux functions, we also reconstruct all magnetic field components at these interfaces. The only difference is that, due to the staggered grid collocation of $\boldsymbol B$, we have to average over pairs of opposing cell interfaces on the way. For the interfaces lying in $x$ direction this means
\begin{align}
B_{x\,i,j,k}^E=&B_{x\,i+\frac{1}{2},j,k}\\ \notag
B_{y\,i,j,k}^E=&\frac{1}{2}\Big(B_{y\,i,j+\frac{1}{2},k}+B_{y\,i,j-\frac{1}{2},k}
+(\delta_x B_y)_{i,j+\frac{1}{2},k}+(\delta_x B_y)_{i,j-\frac{1}{2},k}\Big)\\ \notag
B_{z\,i,j,k}^E=&\frac{1}{2}\Big(B_{z\,i,j,k+\frac{1}{2}}+B_{z\,i,j,k-\frac{1}{2}}
-(\delta_x B_z)_{i,j,k+\frac{1}{2}}-(\delta_x B_z)_{i,j,k-\frac{1}{2}}\Big)\\  \notag
B_{x\,i,j,k}^W=&B_{x\,i-\frac{1}{2},j,k}\\  \notag
B_{y\,i,j,k}^W=&\frac{1}{2}\Big(B_{y\,i,j+\frac{1}{2},k}+B_{y\,i,j-\frac{1}{2},k}
+(\delta_x B_y)_{i,j+\frac{1}{2},k}+(\delta_x B_y)_{i,j-\frac{1}{2},k}\Big)\\  \notag
B_{z\,i,j,k}^W=&\frac{1}{2}\Big(B_{z\,i,j,k+\frac{1}{2}}+B_{z\,i,j,k-\frac{1}{2}}
-(\delta_x B_z)_{i,j,k+\frac{1}{2}}-(\delta_x B_z)_{i,j,k-\frac{1}{2}}\Big)\\  \notag
\end{align}

Note that the $B_x$ component does not get reconstructed, since it is already stored at the needed interface. Again, for the other two directions there are analogue expressions.

Now, we calculate the flux functions $\boldsymbol{\boldsymbol{f}}$ at E,W,N,S,T,B locations by putting the interface values of $\boldsymbol{u}$ and $\boldsymbol B$ into the definition (\ref{def_fluxfunc}). Once we have them, the numerical fluxes in and out of each cell at each interface are calculated utilizing the KNP flux formula:
\begin{align}
\label{knpfluxformula}
\boldsymbol{F}^x_{i+\frac{1}{2},j,k}=&\frac{1}{a^+_{i+\frac{1}{2},j,k}-a^-_{i+\frac{1}{2},j,k}}
\Big[ 
   a^+_{i+\frac{1}{2},j,k}\boldsymbol{f}^x(\boldsymbol{u}^E_{i,j,k},\boldsymbol{B}^E_{i,j,k}) -\\ \notag
   &- a^-_{i+\frac{1}{2},j,k}\boldsymbol{f}^x(\boldsymbol{u}^W_{i+1,j,k},\boldsymbol{B}^W_{i+1,j,k})
   + a^+_{i+\frac{1}{2},j,k} a^-_{i+\frac{1}{2},j,k} \boldsymbol{u}^W_{i+1,j,k}-\boldsymbol{u}^E_{i,j,k})
\Big]\\ \notag
\boldsymbol{F}^y_{i,j+\frac{1}{2},k}=&\frac{1}{b^+_{i,j+\frac{1}{2},k}-b^-_{i,j+\frac{1}{2},k}}
\Big[
   b^+_{i,j+\frac{1}{2},k}\boldsymbol{f}^x(\boldsymbol{u}^N_{i,j,k},\boldsymbol{B}^N_{i,j,k}) -\\ \notag
   &- b^-_{i,j+\frac{1}{2},k}\boldsymbol{f}^x(\boldsymbol{u}^S_{i,j+1,k},\boldsymbol{B}^S_{i,j+1,k})
   + b^+_{i,j+\frac{1}{2},k} b^-_{i,j+\frac{1}{2},k} \boldsymbol{u}^S_{i,j+1,k}-\boldsymbol{u}^N_{i,j,k})
\Big]\\ \notag
\boldsymbol{F}^z_{i,j,k+\frac{1}{2}}=&\frac{1}{c^+_{i,j,k+\frac{1}{2}}-c^-_{i,j,k+\frac{1}{2}}}
\Big[
   c^+_{i,j,k+\frac{1}{2}}\boldsymbol{f}^x(\boldsymbol{u}^T_{i,j,k},\boldsymbol{B}^T_{i,j,k}) -\\ \notag
   &- c^-_{i,j,k+\frac{1}{2}}\boldsymbol{f}^x(\boldsymbol{u}^B_{i,j,k+1},\boldsymbol{B}^B_{i,j,k+1})
   + c^+_{i,j,k+\frac{1}{2}} c^-_{i,j,k+\frac{1}{2}} \boldsymbol{u}^B_{i,j,k+1}-\boldsymbol{u}^T_{i,j,k+1})
\Big]
\end{align}

In these flux formulae, $a^\pm$ denotes the maximum (+) and minimum ($-$) local speed of the hydro density flow at the cell surface in $x$-direction (wavespeed estimate of \citealt{Davis88}):
\begin{align}
a^+_{i+\frac{1}{2},j,k}=\max\{ (v_x+c_f)^W_{i+1,j,k},(v_x+c_f)^E_{i,j,k},0\}\\ \notag
a^-_{i+\frac{1}{2},j,k}=\min\{ (v_x-c_f)^W_{i+1,j,k},(v_x-c_f)^E_{i,j,k},0\}
\end{align}
and $b^\pm$, $c^\pm$ the same for $y$ and $z$, respectively. The expression
\begin{align}
c_f=\sqrt{c_s^2+c_A^2}
\end{align}
is an upper limit for the possible characteristic wave speed in the medium (fast magnetosonic speed), where
\begin{align}
c_s=\sqrt{\frac{\gamma p}{\rho}}
\end{align}
is the sound speed in the medium and
\begin{align}
c_A=\sqrt{\frac{B^2}{\rho}}
\end{align}

is the Alfv\'en speed. All these quantities get calculated on-the-fly using the reconstructed MHD variables and the equations (\ref{eq_of_state}) and (\ref{edens}).

By adding the fluxes through all six cell interfaces, we now have the total flux in and out of the cell:

\begin{align}
\label{total_flux}
\frac{\textrm{d}}{\textrm{d}t}\boldsymbol{u}_{ijk}=&-\frac{\boldsymbol{F}^x_{i+\frac{1}{2},j,k}-\boldsymbol{F}^x_{i-\frac{1}{2},j,k}}{\Delta x}-\frac{\boldsymbol{F}^y_{i,j+\frac{1}{2},k}-\boldsymbol{F}^y_{i-\frac{1}{2},k}}{\Delta y} \\ \notag
&-\frac{\boldsymbol{F}^z_{i,j,k+\frac{1}{2}}-\boldsymbol{F}^z_{i,j,k-\frac{1}{2}}}{\Delta z}
\end{align}
Note that no time discretization has been specified yet. We will later use (\ref{total_flux}) to update $\boldsymbol u_{i,j,k}$ applying a second-order Runge-Kutta scheme for the time integration (see section \ref{subsec:timeintegration}). 

\subsubsection{Constrained transport (CT)}

In order to track the time evolution of the magnetic field $\boldsymbol B$ as well, we want to solve the induction equation (\ref{induction_eq}) in a similar way. But when introducing magnetic fields to such grid algorithms, one is immediately faced with the problem that the solution for $\boldsymbol B$ must comply  \emph{at all times} to the additional condition $\boldsymbol\nabla\cdot\boldsymbol B=0$ down to the highest possible precision. Otherwise magnetic ``sources'' (monopoles) would be introduced that would lead to unphysical results (like forces parallel to the field direction). Physically, $\boldsymbol\nabla\cdot\boldsymbol B$ is a conserved quantity. But this is not the case for numerical calculations -- a nonzero divergence will inevitably build up due to numerical errors, even if $\boldsymbol B$ was divergence-free at the beginning of the simulation. Even worse, numerical nonzero $\boldsymbol\nabla\cdot\boldsymbol B$ usually grows exponentially \citep{Brackbill80}, and the code will crash.

There are a handful of techniques to remedy the situation (see \citealt{Toth00} for a review and comparison study). \citet{Brackbill80} introduced the ``divergence cleaning'' (or ``Hodge Projection'') approach, which solves an extra Poisson's equation to recover $\boldsymbol\nabla\cdot\boldsymbol B=0$ at each time step. But it was found later that the divergence cleaning can introduce substantial amounts of additional spurious structure \citep{Balsara04}. The second method \citep{Powell99,Dedner02} extends the MHD equations to produce an additional divergence wave, which then advects the divergence out of the domain. This generally works; however, in cosmological simulations we always work with periodic boundary conditions. Thus, a wave cannot leave the domain, and this method is not applicable.

In \amiga, we use the arguably most elegant solution, the \emph{constrained transport} (CT) method by \citet{Evans88}. In this method, the components of $\boldsymbol B$ are arranged in a way that ensures  $\boldsymbol\nabla\cdot\boldsymbol B=0$ by definition. This is the reason why we introduced the staggered grid.

Another issue with incorporating the induction law in the chosen finite-volume scheme is that the conserved quantity, i.e. the magnetic flux, is defined on a surface rather than on a volume (such as density). This leads to the fact that, as opposed to equations (\ref{density_eq})--(\ref{energy_eq}), the induction equation (\ref{induction_eq}), albeit it is a conservation law, contains the curl operator $\boldsymbol\nabla\times$ instead of the divergence operator $\boldsymbol\nabla\cdot\;$. Here, it is possible to apply a trick that still allows to use the same numerical scheme for the magnetic field as for the hydro variables $\boldsymbol u$. If we write $\boldsymbol E=-\boldsymbol v\times\boldsymbol B$, then the induction equation becomes

\begin{align}
\label{induction_eq_sourcefree}
\frac{\partial \boldsymbol B}{\partial t} + \boldsymbol\nabla\times\boldsymbol E = \boldsymbol S_B
\end{align}
with the magnetic Hubble source term $\boldsymbol S_B = \frac{1}{2} \mathcal H \boldsymbol B$ (in the non-cosmological case it is zero). Now this equation can be transformed into divergence form using an antisymmetric flux tensor, i.e.

\begin{align}
\label{fluxtensor}
\frac{\partial \boldsymbol B}{\partial t} + \boldsymbol\nabla\cdot
\left(
   \begin{array}{ccc}
   0&E_z&-E_y\\ 
   -E_z&0&-E_x \\
   E_y&-E_x&0
   \end{array}
\right)= \boldsymbol S_B
\end{align}
which is formally analogous to (\ref{general_conserv}); but instead of flux functions (\ref{def_fluxfunc}), we use the components of the antisymmetric flux tensor. It is essentially just a ``resorting'' of the vector components to make the curl appear formally as a divergence. With this, it is possible to construct numerical fluxes $\boldsymbol{E}$ using the same KNP flux formula as before:

\begin{align}
\label{knpfluxmhd}
\boldsymbol{E}^x_{i+\frac{1}{2},j,k}=&\frac{1}{a^+_{i+\frac{1}{2},j,k}-a^-_{i+\frac{1}{2},j,k}}
\Bigg[ 
   a^+_{i+\frac{1}{2},j,k}\begin{pmatrix}0\\-E_z\\E_y\end{pmatrix}^E_{i,j,k} -\\ \notag
   &- a^-_{i+\frac{1}{2},j,k}\begin{pmatrix}0\\-E_z\\E_y\end{pmatrix}^W_{i+1,j,k}
   + a^+_{i+\frac{1}{2},j,k} a^-_{i+\frac{1}{2},j,k}( \boldsymbol B^W_{i+1,j,k}-\boldsymbol B^E_{i,j,k})
\Bigg] \displaybreak[0] \\ \notag
\boldsymbol{E}^y_{i,j+\frac{1}{2},k}=&\frac{1}{b^+_{i,j+\frac{1}{2},k}-b^-_{i,j+\frac{1}{2},k}}
\Bigg[ 
   b^+_{i,j+\frac{1}{2},k}\begin{pmatrix}E_z\\0\\-E_x\end{pmatrix}^N_{i,j,k} -\\ \notag
   &- b^-_{i,j+\frac{1}{2},k}\begin{pmatrix}E_z\\0\\-E_x\end{pmatrix}^S_{i,j+1,k}
   + b^+_{i,j+\frac{1}{2},k} b^-_{i,j+\frac{1}{2},k}( \boldsymbol B^S_{i,j+1,k}-\boldsymbol B^N_{i,j,k})
\Bigg] \displaybreak[0] \\ \notag
\boldsymbol{E}^z_{i,j,k+\frac{1}{2}}=&\frac{1}{c^+_{i,j,k+\frac{1}{2}}-c^-_{i,j,k+\frac{1}{2}}}
\Bigg[ 
   c^+_{i,j,k+\frac{1}{2}}\begin{pmatrix}-E_y\\E_x\\0\end{pmatrix}^T_{i,j,k} -\\ \notag
   &- c^-_{i,j,k+\frac{1}{2}}\begin{pmatrix}-E_y\\E_x\\0\end{pmatrix}^B_{i,j,k+1}
   + c^+_{i,j,k+\frac{1}{2}} c^-_{i,j,k+\frac{1}{2}}( \boldsymbol B^B_{i,j,k+1}-\boldsymbol B^T_{i,j,k})
\Bigg]
\end{align}

Now, to ensure $\boldsymbol\nabla\cdot\boldsymbol B=0$, the constrained transport method enters. The idea is to discretize the magnetic field and the magnetic fluxes in such a way that the $\boldsymbol\nabla\cdot\boldsymbol B=0$ condition follows by definition of the scheme and is therefore conserved down to machine precision.

\begin{figure}
\begin{center}
\begin{minipage}{0.40\textwidth}
        \epsfig{file=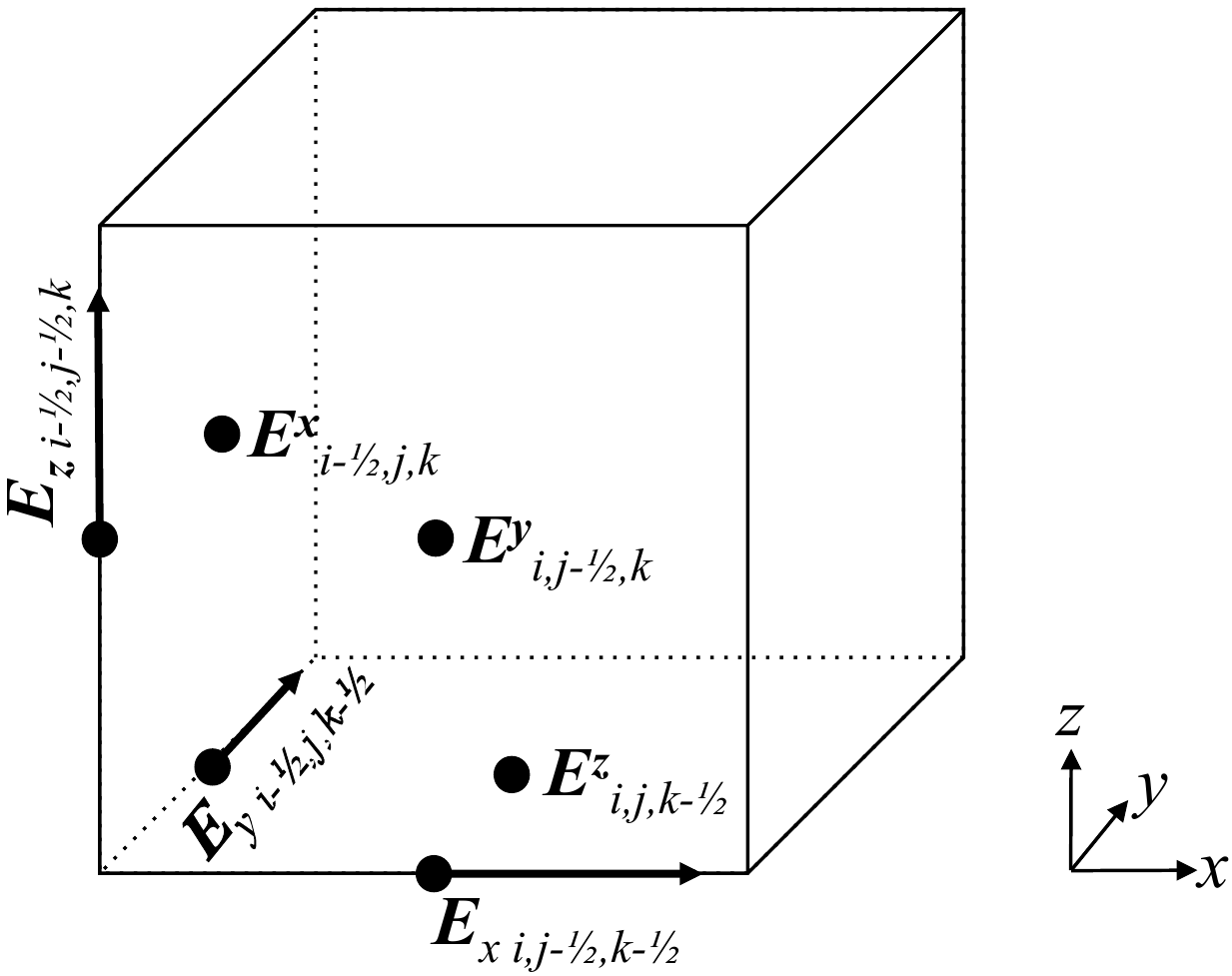, width=1\textwidth, angle=0}
\end{minipage}
\end{center}
\caption{Face-centered and edge-centered $\boldsymbol E$ fields in the constrained transport scheme.
\label{fig:ctcell}}
\end{figure}

From the face-centered fluxes (\ref{knpfluxmhd}) we calculate edge-centered fluxes. An easy way to calculate $\boldsymbol E$ on a cell edge is averaging over the four interfaces touching this edge (see Figure \ref{fig:ctcell}):

\begin{align}
\label{CT_ziegler}
E_{x\,i,j-\frac{1}{2},k-\frac{1}{2}}=
\frac{1}{4}(-\boldsymbol E^y_{z\,i,j-\frac{1}{2},k}
-\boldsymbol E^y_{z\,i,j-\frac{1}{2},k-1}\\ \notag
+\boldsymbol E^z_{y\,i,j,k-\frac{1}{2}}
+\boldsymbol E^z_{y\,i,j-1,k-\frac{1}{2}}) \\ \notag
E_{y\,i-\frac{1}{2},j,k-\frac{1}{2}}=
\frac{1}{4}(-\boldsymbol E^x_{z\,i-\frac{1}{2},j,k}
-\boldsymbol E^x_{z\,i-\frac{1}{2},j,k-1}\\ \notag
+\boldsymbol E^z_{x\,i,j,k-\frac{1}{2}}
+\boldsymbol E^z_{x\,i-1,j,k-\frac{1}{2}})\\ \notag
E_{z\,i-\frac{1}{2},j-\frac{1}{2},k}=
\frac{1}{4}(-\boldsymbol E^x_{y\,i-\frac{1}{2},j,k}
-\boldsymbol E^x_{y,i-\frac{1}{2},j,k-1}\\ \notag
+\boldsymbol E^y_{x\,i,j-\frac{1}{2},k}
+\boldsymbol E^y_{x\,i-1,j-\frac{1}{2},k})\\ \notag
\end{align}

The other nine edges bordering cell $i,j,k$ are obtained in the same way with the according indices. It has been pointed out recently by \citet{Gardiner05,Gardiner08} that this is actually not the best way to construct edge-centered fluxes and that for certain cases, a reconstruction algorithm with proper upwinding gives better results than the simple averaging. However, for the tests and simulations presented here, this has no relevant effects. If it becomes necessary in the future to improve the algorithm, another scheme like the one by \citet{Gardiner05,Gardiner08} may be implemented by simply adjusting equation~(\ref{CT_ziegler}) accordingly. Alternatively, a numerical dissipation term can be introduced in the induction equation to smear out any possible numerical noise.

Using the edge-centered fluxes, we get the temporal change of the staggered magnetic field components, in analogy to the hydro flux (\ref{total_flux}):

\begin{align}
\label{total_fluxmhd}
\frac{d}{dt}B_{x\,i-\frac{1}{2},j,k}=-\frac{E_{z\,i-\frac{1}{2},j+\frac{1}{2},k}-E_{z\,i-\frac{1}{2},j-\frac{1}{2},k}}{\Delta y}\\ \notag
+\frac{E_{y\,i-\frac{1}{2},j,k+\frac{1}{2}}-E_{y\,i-\frac{1}{2},j,k-\frac{1}{2}}}{\Delta z}\\ \notag
\frac{d}{dt}B_{y\,i,j-\frac{1}{2},k}=-\frac{E_{z\,i+\frac{1}{2},j-\frac{1}{2},k}-E_{z\,i-\frac{1}{2},j-\frac{1}{2},k}}{\Delta x}\\ \notag
+\frac{E_{x\,i,j-\frac{1}{2},k+\frac{1}{2}}-E_{x\,i,j-\frac{1}{2},k-\frac{1}{2}}}{\Delta z}\\ \notag
\frac{d}{dt}B_{z\,i+\frac{1}{2},j,k-\frac{1}{2}}=-\frac{E_{y\,i+\frac{1}{2},j,k-\frac{1}{2}}-E_{y\,i-\frac{1}{2},j,k-\frac{1}{2}}}{\Delta x}\\ \notag
+\frac{E_{x\,i,j+\frac{1}{2},k-\frac{1}{2}}-E_{x\,i,j-\frac{1}{2},k-\frac{1}{2}}}{\Delta y}
\end{align}

By writing out  $\boldsymbol \nabla\cdot\boldsymbol B$ with these discretizations one can immediately see that $\textrm{d}(\boldsymbol\nabla\cdot\boldsymbol B)/\textrm{d}t = 0$ by definition. Therefore, with compatible initial conditions, $\boldsymbol\nabla\cdot\boldsymbol B=0$ is conserved at all times.

\subsubsection{Time integration}
\label{subsec:timeintegration}

In order to calculate the temporal changes of the MHD variables, we discretize the equations (\ref{total_flux}) and (\ref{total_fluxmhd}) in time with a standard second-order accurate two-step Runge-Kutta method. At a given time $t$, we have $\boldsymbol u^t$ and $\boldsymbol B^t$ stored as described before. First, we estimate the appropriate timestep $\Delta t$ by using the usual timestep criteria. The time step $\Delta t$ should not exceed the actual age of the universe,

\begin{align}
\Delta t \le \frac{1}{\mathcal H}\;\;\; ,
\end{align}
the fastest dark matter particle in the box should not travel farther than some fraction $\epsilon_1$ of one grid cell during one timestep,
\begin{align}
\Delta t \le \frac{\epsilon_1 \; \Delta x}{v_{DM,max}}\;\;\; ,
\end{align}
and the same must hold for the fastest baryon flow speed encountered in the medium,
\begin{align}
\label{cflcondition}
\Delta t \le \frac{\epsilon_2 \; \Delta x}{v_{max}}\;\;\; ,
\end{align}
where $\epsilon_2$, the so-called CFL number, should always be $\le 0.5$  (CFL criterion).
 Now, denoting the right-hand side of equation (\ref{total_flux}) as $\boldsymbol F(\boldsymbol u, \boldsymbol B)$ and the right-hand side of equation (\ref{total_fluxmhd}) as $\boldsymbol E(\boldsymbol u, \boldsymbol B)$, we perform a predictor timestep
\begin{align}
\boldsymbol u^{t+\Delta t*}&=\boldsymbol u^t+\Delta t \cdot\boldsymbol F(\boldsymbol u^t, \boldsymbol B^t)\\ \notag
\boldsymbol B^{t+\Delta t*}&=\boldsymbol B^t+\Delta t \cdot\boldsymbol E(\boldsymbol u^t, \boldsymbol B^t)
\end{align}
These predictor values are used to calculate timestep-centered values
\begin{align}
\label{u_tmp}
\boldsymbol u^{t+\frac{\Delta t}{2}*}&=\frac{1}{2}(\boldsymbol u^t+\boldsymbol u^{t+\Delta t*})\\ \notag
\boldsymbol B^{t+\frac{\Delta t}{2}*}&=\frac{1}{2}(\boldsymbol B^t+\boldsymbol B^{t+\Delta t*})
\end{align}
and in a final step, $\boldsymbol u$ and $\boldsymbol B$ get stepped forward in time using these timestep-centered values:
\begin{align}
\boldsymbol u^{t+\Delta t}=&\boldsymbol u^{t+\frac{\Delta t}{2}*}+\frac{\Delta t}{2}\cdot\boldsymbol F(\boldsymbol u^{t+\Delta t*}, \boldsymbol B^{t+\Delta t*})+\Delta t \cdot \boldsymbol S_u^{t+\frac{\Delta t}{2}*}\\ \notag
\boldsymbol B^{t+\Delta t}=&\boldsymbol B^{t+\frac{\Delta t}{2}*}+\frac{\Delta t}{2}\cdot\boldsymbol E(\boldsymbol u^{t+\Delta t*}, \boldsymbol B^{t+\Delta t*})+\Delta t \cdot \boldsymbol S_B^{t+\frac{\Delta t}{2}*}\\ \notag
\end{align}
where $\boldsymbol S_u$ are the timestep-averaged hydro source terms (\ref{u_source}) and $\boldsymbol S_B=\mathcal H\boldsymbol B/2$ is the magnetic Hubble term from equation (\ref{induction_eq}).

\begin{figure}
\begin{center}
\begin{minipage}{0.475\textwidth}
        \epsfig{file=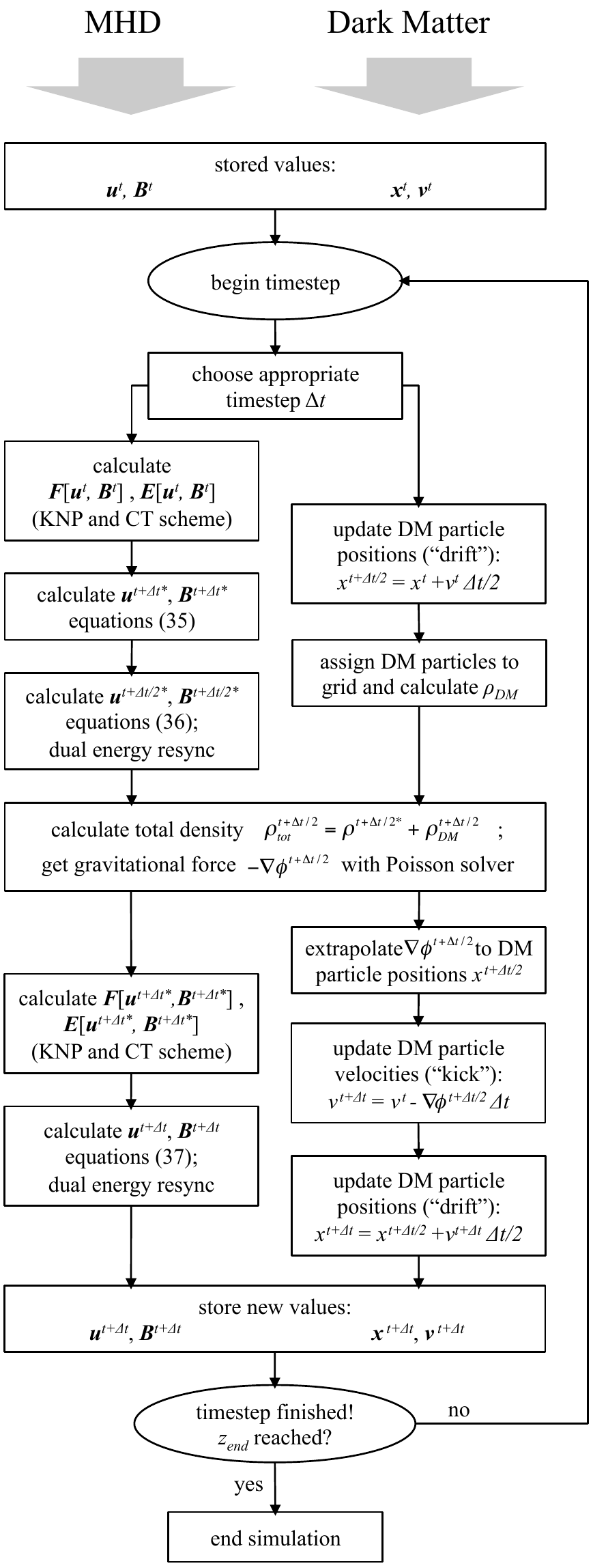, width=1\textwidth, angle=0}
\end{minipage}
\end{center}
\caption{Flowchart of \amiga's timestepping on a regular grid. The stored values are dark matter positions $\boldsymbol x$ and velocities $\boldsymbol v$, the hydrodynamical quantities $\boldsymbol u$ (cell-centered values) and the magnetic field $\boldsymbol B$ (staggered face-centered values).
\label{fig:flowchart}}
\end{figure}

It is important to point out that we must use a different time integration scheme here than the one in the $N$-body part. While leapfrog-based integrators like the one used by the $N$-body solver are well suited for Hamiltonian-type equations of motion and in particular the $N$-body problem, they are unstable for hyperbolic conservation laws like the MHD equations. However, the time integration schemes in the $N$-body solver and the MHD solver are connected only through the gravitational potential $\phi$. For both time integrators, the \emph{timestep-averaged} gravitational potential $\phi^{t+\frac{\Delta t}{2}}$ is needed, so we can compute that from the time-averaged total density $\rho_{tot}^{t+\frac{\Delta t}{2}}$ (the sum of baryon and dark matter density). Figure \ref{fig:flowchart} shows a flowchart of a full \amiga\ timestep, illustrating how the $N$-body-solver and the MHD algorithm are interconnected. 

The code is completely modular, i.e. it is possible to run a pure $N$-body simulation, a pure hydrodynamic simulation, an MHD simulation or a combination of everything. For non-cosmological runs like the test cases presented before, it is possible to set $a=1, \dot a=0$, and the supercomoving MHD equations reduce to the ordinary MHD equations in proper physical coordinates. The gravity solver and the periodic boundary conditions can also be modified or disabled.

\subsubsection{Supersonic flows and the dual energy formalism}
\label{dualenergy}

When including gas physics in cosmological simulations, due to the extreme gravitational forces the gas flows can be accelerated to highly supersonic speeds, reaching Mach numbers of 100 and more. While the shocks and discontinuities that are created by such flows can be captured very well by the KNP solver, they are also followed by highly supersonic bulk flows of cold gas. A serious numerical problem occurs when trying to describe such flows with ideal MHD equations.

At different places in the code, the thermal energy density $\rho\varepsilon$ and pressure $p$ have to be calculated. Normally, this happens through equations (\ref{eq_of_state}) and (\ref{edens}). The problem lies in the fact that in such cold, highly supersonic bulk flows, the value of $\rho \varepsilon$ will become several orders of magnitude smaller than the total energy density $\rho E$. Expression (\ref{edens}) then contains a small difference of large numbers, leading to wrong results due to limited floating point precision. The thermal pressure and therefore the gas temperature cannot be tracked accurately anymore.

To remedy this situation, \citet{Ryu93} proposed to introduce the modified entropy as an additional equation from which the thermal energy could be calculated. Alternatively, \citet{Bryan95} use an equation for the thermal energy itself (which is a bit problematic because it is not a conservation law). We follow the \citet{Ryu93} method and define the modified entropy as
\begin{align}
\label{entropy_def}
S=\frac{p}{\rho^{\gamma-1}}\;\;\;.
\end{align}
In supercomoving coordinates, the supercomoving modified entropy follows the conservation law 
\begin{align}
\label{entropy_eq}
\frac{\partial S}{\partial t}+\boldsymbol\nabla\cdot(S\boldsymbol v)=- \mathcal H S (3\gamma -5)
\end{align}
(see appendix for the derivation), where the right-hand side equals zero for $\gamma = 5/3$.

For the whole simulation, we solve this equation alongside equations (\ref{density_eq}) through (\ref{energy_eq}) with the KNP solver. Now, whenever the thermal energy cannot be calculated accurately from the usual set of equations (\ref{density_eq}), (\ref{momentum_eq}) and (\ref{energy_eq}) -- the ``E system'' -- , we use the set of equations (\ref{density_eq}), (\ref{momentum_eq}) and (\ref{entropy_eq}) -- the ``S system'' -- instead. In this system, the pressure and thermal energy are calculated as follows:
\begin{align}
\label{S_edens}
p=S\rho^{\gamma-1}\; ;    \;\;     \rho\varepsilon=\frac{S\rho^{\gamma-1}}{\gamma-1}
\end{align}
After each timestep, the two systems have to be resynchronized: if the S system was used, the total energy has to be updated to be consistent with the new internal energy; if the E system was used, the entropy has to be updated according to equation (\ref{entropy_def}).

The crucial step here is how to determine when to use the entropy for the calculation. A possible choice is to do this whenever the ratio of $\rho\varepsilon$ to $\rho E$ gets smaller than some threshold parameter (e.g. in \citealt{Collins09}):
\begin{align}
\label{eta1}
\frac{\rho E-\rho v^2/2-B^2/2}{\rho E}<\eta_1
\end{align}
This works very well for most cases. However, in cosmological simulations sometimes another situation occurs when \emph{all} energy components are near zero numerically, for example in the low-density regions between the shocks in the double pancake test (see section \ref{subsec:doublepancake}). The condition (\ref{eta1}) is false, nevertheless the pressure can not be tracked accurately. In order to deal with this issue, we propose a new approach: Instead of using the S system only in cases where~(\ref{eta1}) is true, we reverse the original condition and use the S system always, given that the entropy is conserved (that is, outside of shocks). Whether we are in a shock or not gets estimated by an additional criterion, checking for steep pressure gradients:
\begin{align}
\label{eta2}
\frac{|\boldsymbol\nabla p|}{p}<\eta_2
\end{align}
Whenever either (\ref{eta1}) or (\ref{eta2}) is true in a grid cell, we calculate the thermal energy using the S system. This new method gives accurate thermal quantities not only in strong shocks, but also in very cold low-density regions. Of course, since technically the S system abandons strict energy conservation in favor of accurately tracking the temperature, we must make sure that the use of the S system does not have a dynamical effect on the other hydrodynamic variables by choosing the parameters low enough. For the cosmological MHD simulations presented here, we used $\eta_1=0.001$ and $\eta_2=0.3$.

The cell-averaged value of the magnetic energy density $B^2/2$ is required here for compatibility with the other energy terms. It is calculated by averaging over the opposing pairs of face-centered $\boldsymbol B$ components that enclose the cell:
\begin{align}
\label{cellcenteredB2}
\left (\frac{B^2}{2}\right)_{i,j,k}&=\frac{1}{8}\Big[ (B_{x\;i+\frac{1}{2},j,k}+B_{x\;i-\frac{1}{2},j,k})^2 \\ \notag 
&+(B_{y\;i,j+\frac{1}{2},k}+B_{y\;i,j-\frac{1}{2},k})^2 +(B_{z\;i,j,k+\frac{1}{2}}+B_{z\;i,j,k-\frac{1}{2}})^2 \Big]
\end{align}

\subsection{Code testing}
\label{sec:codetesting}

To verify that the code is working correctly we applied it to a set of standard test problems. The $N$-body solver of \amiga\ comes from its predecessor \mlapm\ \citep{Knebe01}. It has been thoroughly tested therein and successfully used for cosmological dark matter simulations (e.g. \citealt{Gill04a,Gill04b, Gill05, Warnick06, Warnick08}). Therefore we can concentrate here on testing the newly implemented MHD solver and its interplay with the gravity solver.


The hydrodynamic part of the solver is applied on a 1D test, the Sod shock tube \citep{Sod78}, and a 3D test, the Sedov-Taylor blast wave \citep{Sedov59}. To verify the MHD solver and the CT scheme we use the Brio-Wu problem \citep{BrioWu88} and the Orszag-Tang vortex \citep{OrszagTang79}. Then, combining MHD with the gravity solver and the cosmological expansion, we present the double pancake test of \citet{Bryan95}, which also serves as a stringent test on the dual energy algorithm. 

The computational domain for all tests is $x,y,z\in [0,1]$, conforming with the internal code units. All numerical runs up to $N=256$ cells of box length have been performed on the full three-dimensional $N^3$ box, even for 1D test problems, to test the code under more realistic circumstances. For higher resolutions we used a reduced 1D box to save computing time. It turned out that both recover the exact same result. Furthermore, in the case of pure hydrodynamic tests with no magnetic field, full MHD runs with the initial $\boldsymbol B$ set to zero recover the exact same result as purely hydrodynamic runs.


\subsubsection{Sod Shock tube}

\begin{figure*}
\begin{center}
\begin{minipage}{0.84\textwidth}
        \epsfig{file=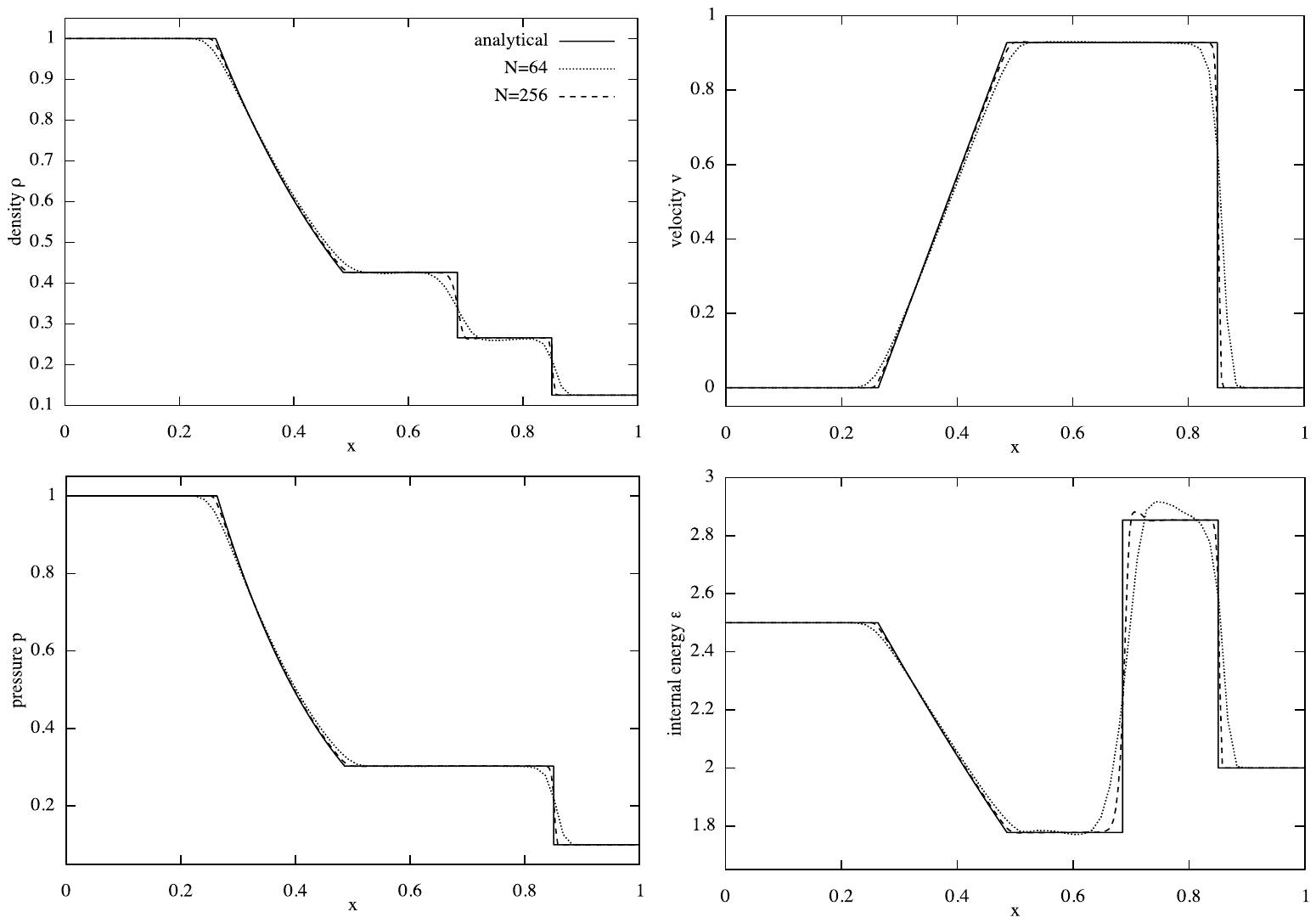, width=1\textwidth, angle=0}
\end{minipage}
\end{center}
\caption{Numerical solution of the shock tube test with different resolutions. The main shock at $x=0.85$ is always between three and four grid cells wide. The reference solution was generated with an exact Riemann solver algorithm based on \citet{Toro99}.
\label{fig:shocktube}}
\end{figure*}

For the Sod shock tube test, the simulation box is divided in two halves with constant initial states separated by a barrier between them that is removed at $t=0$. This generates a strong shock wave moving to the lower density region, a sound wave (rarefaction) in the opposite direction and a contact discontinuity. This simultaneous presence of different phases makes the shock tube an excellent method to check how well a code handles strong shocks. We chose the same initial conditions as in the original paper of \citet{Sod78}. The left and right initial states at $t=0$ are:
\begin{align*}
&\rho_L=1\; ; \;\;&\rho_R=0.125\\
&p_L=1\; ; \;\;&p_R=0.1
\end{align*}
The initial velocity is zero everywhere, the polytropic index is $\gamma=1.4$. The boundary conditions are non-periodic and the system is evolved until $t_{end}=0.2$. By then the main shock will be located at $x=0.85$. Figure \ref{fig:shocktube} shows the results perpendicular to the shock plane.

In the code comparison suite of \citet{Tasker08}, this test is applied to other astrophysical codes. In direct comparison, \amiga\ handles the problem very well. The main shock is between three and four cells wide, an accuracy comparable to PPM grid codes. All features of the analytical solution are recovered very accurately without oscillations or other numerical artefacts, except for a slight overshoot in the internal energy at the contact discontinuity. This is a common feature in grid codes and quickly disappears with higher resolution.

The analytical reference solution for this problem was generated with an exact Riemann solver algorithm based on \citet{Toro99}, using a resolution of $N=10^4$.


\begin{figure*}
\begin{center}
\begin{minipage}{0.84\textwidth}
        \epsfig{file=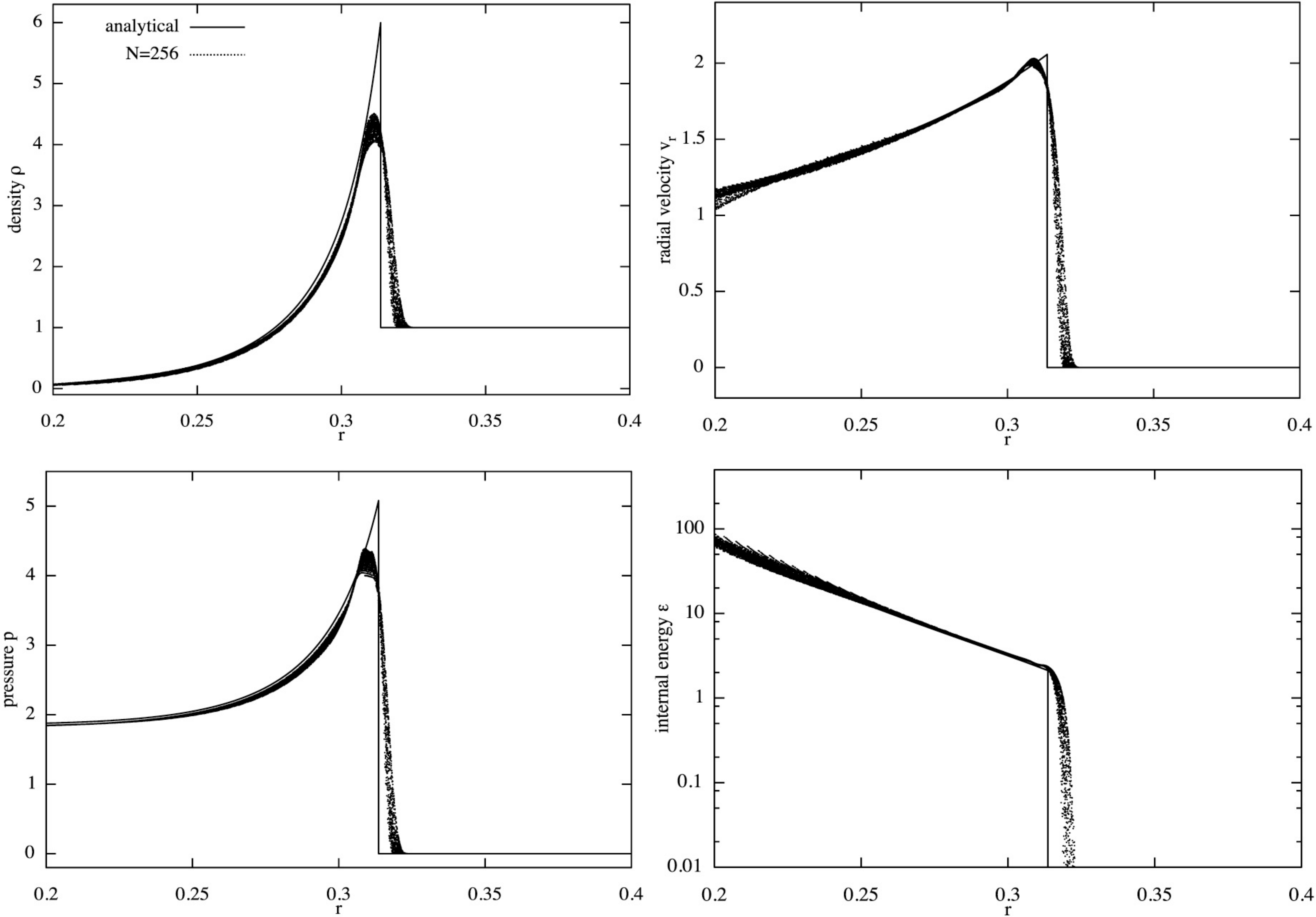, width=1\textwidth, angle=0}
\end{minipage}
\end{center}
\caption{Numerical solution of the Sedov blast wave. The dots correspond to cell values of the \amiga\ run at $t=0.508$ with a resolution of $256^3$; the solid line is the reference solution computed from the analytical formulae given in \citet{Sedov59}.
\label{fig:sedov}}
\end{figure*}

\begin{figure*}
\begin{center}
\begin{minipage}{1.0\textwidth}
        \epsfig{file=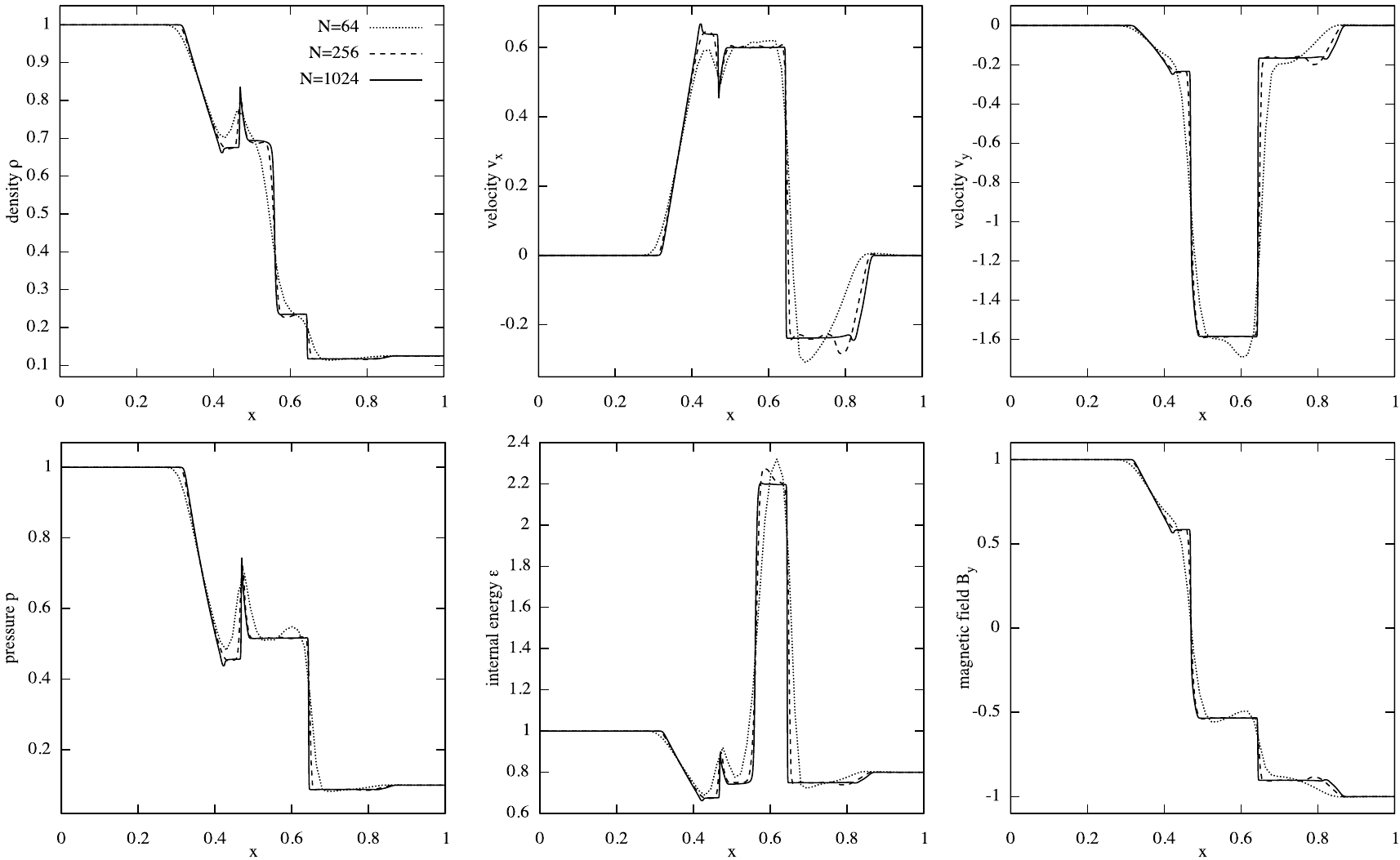, width=1\textwidth, angle=0}
\end{minipage}
\end{center}
\caption{Numerical solution of the Brio-Wu problem. Since no analytical solution is known, a high-resolution run with $N=1024$ serves as the reference solution. This test can be found in e.g. \citet{Ryu95}. There is an overshoot present in the x-direction velocity that disappears only at higher resolutions; otherwise, the results compare and converge extremely well.
\label{fig:briowu}}
\end{figure*}

\begin{figure*}
\begin{center}
\begin{minipage}{1.0\textwidth}
        \epsfig{file=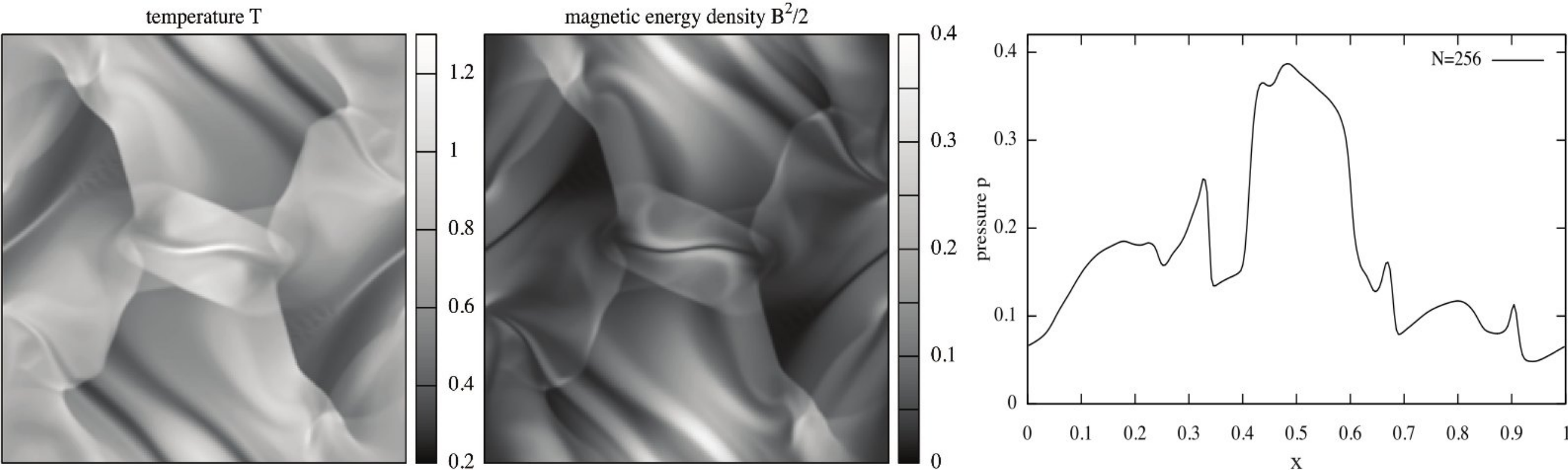, width=1\textwidth, angle=0}
\end{minipage}
\end{center}
\caption{Numerical solution of the Orszag-Tang vortex at $t=0.5$ with $N=256$ resolution. From left to right: temperature and magnetic energy density distribution in the $x$-$y$ plane; gas pressure along a cut at $y=0.4277$.\label{fig:orszagtang}}
\end{figure*}



\subsubsection{Sedov Blast wave}

The Sedov blast wave test is performed by injecting a large amount of thermal energy in a small, point-like region with uniform cold gas around it. This causes a strong explosion with a spherical shock front propagating outwards. The test is particularly useful to check if spherical symmetry is preserved by the code. It is important that on a cubic grid, shock fronts that are aligned parallel to the grid are resolved the same way as those moving at an oblique angle, because otherwise we would introduce an artificial anisotropy. Also, the shock front is very narrow and thus numerically challenging to resolve. As a reference we use the known self-similar analytical solution \citep{Sedov59}.

For this test, the gas is at rest with $\rho=1$ and $v=0$ everywhere. We inject the energy $E_0=1$ in a spherical region of radius 3.5 cells in the centre of the simulation box. Then, the initial pressure equals
\begin{align*} 
p = \;\;
\begin{cases} 
\;\;\dfrac{3(\gamma -1)E_0}{4\pi r^3} & \text{if } r < 3.5\; \Delta x \\[1em]
\;\;10^{-5} & \text{else} 
\end{cases}
\end{align*}

We use $\gamma=7/5$ and evolve the blast wave until $t=0.0508$. The shock is then located at $r=0.314$. According to \citet{Tasker08}, the $r=3.5$ cells sphere is a good approximation of a point-like energy source, if we use a uniform grid with $N=256$ or higher; so we use exactly this resolution. The results are compared with the analytical solution in figure~\ref{fig:sedov}.

The code conserves spherical symmetry and recovers the analytical solution well. The shock front is smoothed over a width of approx. four grid cells, so there is not much broadening due to a shock propagation on different angles with respect to the grid. The anisotropic scatter is not larger than one grid cell. The peak amplitude is somewhat lower than the analytical solution, but still very well compared with other codes \citep{Ricker00,Tasker08}. The lowering is partly due to the fact that we use a finite spherical region instead of a really point-like source, which is just impossible with a grid code. Some codes also suffer from other problems: the shock position is sometimes underestimated by as much as 4\%, for example \textsc{Enzo} (\textsc{Zeus}) in the \citet{Tasker08} code comparison, since the initial energy lies in a region made of cubical cells and is therefore not exactly spherically symmetric. It can result in a deformed shockwave lagging behind the analytical solution, and the position will be wrong. However, \amiga\ does not suffer from such problems, even if no technique is applied to make the start region more spherical (e.g. Gaussian smoothing or some other weighted distribution), and always recovers the correct shock front position.

For both hydrodynamic tests in general, we find that the numerical accuracy of \amiga's hydrodynamic shock capturing is on par with the most popular astrophysical grid codes used today.


\subsubsection{Brio-Wu problem}

Now we want to test whether the MHD equations are implemented correctly into the solver. We use the test of  \citet{BrioWu88}, which is one-dimensional, so the constrained transport reduces to a simple advection (we will move on to a multi-dimensional test afterwards). The Brio-Wu test is very similar to the Sod shock tube, with left and right initial states and a Riemann discontinuity in between; but in addition it features a magnetic field that has components both parallel and perpendicular to the shock plane, interacting with the shock and the different discontinuities. We use it to check the correct implementation of MHD equations and MHD shock capturing in one dimension, before continuing with multidimensional and cosmological tests. The initial conditions for this test are:
\begin{align*}
\rho_L&=1\; ; &\rho_R&=0.125\\
p_L&=1\; ; \;\;&p_R&=0.1\\
\boldsymbol B_L&=\begin{pmatrix}  0.75\\  1\\  0  \end{pmatrix} \;\;\;\;
&\boldsymbol B_R&=   \begin{pmatrix}  0.75\\  -1\\  0   \end{pmatrix}
\end{align*}
This leads to the propagation of all seven MHD waves (2 shocks, 2 Alfv\'en waves, 2 slow magnetosonic waves and a contact discontinuity) to travel through the box. For these initial conditions, two of the waves will have almost the same speed and interfere with each other, causing the overshoots typical for this test.

Again, the velocity is zero everywhere, but this time we use $\gamma=2$. The system is evolved until $t=0.1$. Since there is no analytical solution known for this problem, we use a high-resolution run with $N=1024$ as a reference. For comparison, this test can also be found e.g. in \citet{Ryu95}. The numerical results obtained with \amiga\ are shown in figure \ref{fig:briowu}. In the pre-shock region, there is an additional overshoot in the x-direction velocity that disappears only at higher resolutions; otherwise, the results are quite accurate and show a good convergence with higher resolution. We acknowledge that the MHD equations are implemented correctly and the shock capturing is accurately handled by the code in the full MHD case. 


\subsubsection{Orszag-Tang vortex}

The next task is to check the multidimensional MHD behaviour: the correct implementation of the constrained transport algorithm and the conservation of $\boldsymbol\nabla\cdot\boldsymbol B=0$. One of the most popular benchmark tests for that purpose is the Orszag-Tang vortex \citep{OrszagTang79}. This 2D test features an initially smooth flow that quickly develops MHD shocks and shock-shock interactions, and eventually breaks down into supersonic MHD turbulence. The initial conditions for this test are:

\begin{align}
&\rho=\frac{25}{36\pi} \; ; 
&p &= \frac{5}{12 \pi} \\ \notag
&\boldsymbol v= 
   \begin{pmatrix}
  -\sin (2\pi y)\\
  \sin (2 \pi x)\\
  0
   \end{pmatrix}  \; ; 
&\boldsymbol{B}&= \frac{1}{\sqrt{4\pi}}
   \begin{pmatrix}
  -\sin (2\pi y)\\
  \sin (4 \pi x)\\
  0
   \end{pmatrix} 
\end{align}

We use periodic boundary conditions and $\gamma=5/3$, leading to $c_s=\sqrt{\gamma p/\rho}=1$ everywhere. Note that here, as for any MHD test, the initial magnetic field must be chosen so that it is divergence-free. The system is then evolved until $t=0.5\;$.

The numerical results are shown in figure \ref{fig:orszagtang}: maps of the temperature and magnetic field energy density in the computational plane, and the gas pressure along a cut at $y=0.4277$. \amiga\ recovers the characteristic, complex shape of the solution in great detail, including the thin thread-like structure in the middle of the box. Magnetic field components are tracked correctly, and most importantly, $\boldsymbol\nabla\cdot\boldsymbol B$ equals machine zero at all times and positions. For comparison, the same test performed on other MHD codes can be found e.g. in \citet{Ryu98}, \citet{Fromang06} (grid codes) and \citet{Borve06}, \citet{Dolag09} (SPH codes).


\begin{figure*}
\begin{center}
\begin{minipage}{1.0\textwidth}
        \epsfig{file=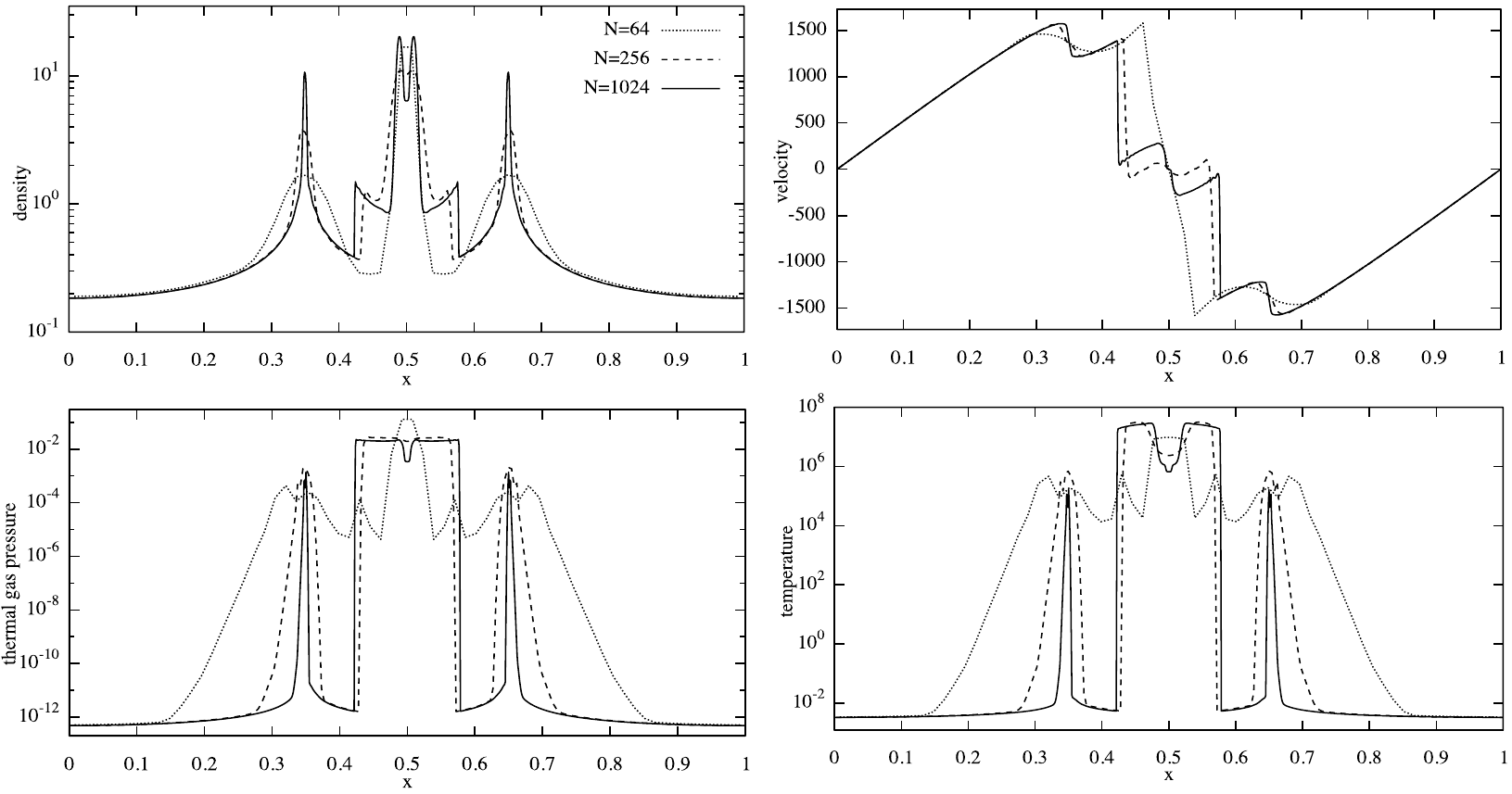, width=1\textwidth, angle=0}
\end{minipage}
\end{center}
\caption{The double pancake problem at $z=0$ with different resolutions (in supercomoving coordinates and code units; velocity in km/s). For this test we set $B=0$ and use the dual energy formalism as described in subsection \ref{dualenergy}. The boxsize is 64 Mpc/h.
\label{fig:double_pancake}}
\end{figure*}

\begin{figure*}
\begin{center}
\begin{minipage}{1.0\textwidth}
        \epsfig{file=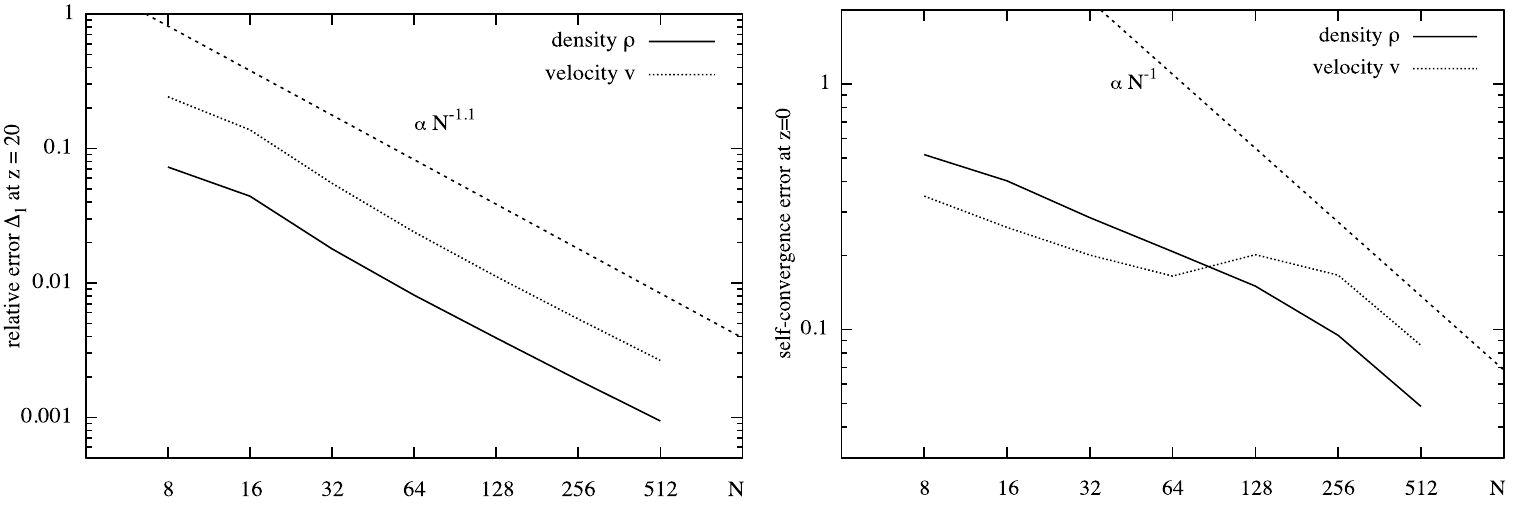, width=1\textwidth, angle=0}
\end{minipage}
\end{center}
\caption{Convergence check for the double pancake problem. The left plot is the relative error in the linear regime (at redshift $z=20$). The right plot is the self-convergence error (definition see text) for the final solution at $z=0$. Since the evolution is highly non-linear at this time and features singularities, the convergence rate is quite low at poorer resolutions (around $N^{-0.5}$).
\label{fig:pancake_error}}
\end{figure*}


\begin{figure*}
\begin{center}
\begin{minipage}{1\textwidth}
        \epsfig{file=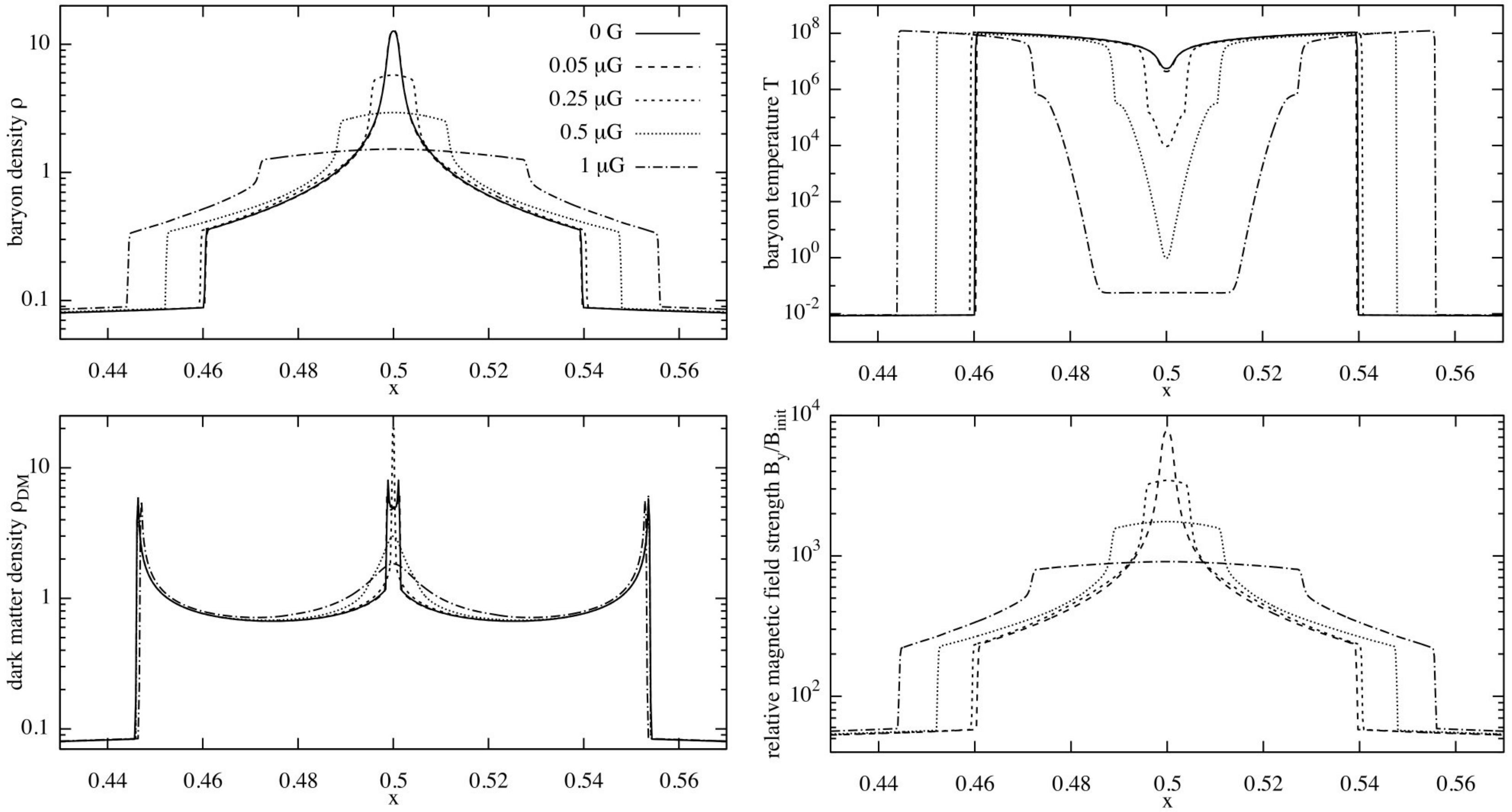, width=1\textwidth, angle=0}
\end{minipage}
\end{center}
\caption{Cosmological MHD pancake at $z=0$ with perpendicular magnetic field and dark matter for different initial field strengths $B_{init}$. The test is run in a one-dimensional box with $N=4096$ resolution for the hydrodynamical grid and 4 dark matter particles per grid cell.
\label{fig:MHDpancake}}
\end{figure*}


\subsubsection{Double pancake test}
\label{subsec:doublepancake}
The cosmological pancake formation \citep{Zeldovich70} is a very popular test for cosmological hydrocodes, because it combines all of the essential physics (hydrodynamics, cosmological expansion and gravity) and is a very stringent test due to the strong shocks and non-linearities present after the caustic formation. Also, since it describes the evolution of a periodic, sinusoidal density perturbation with a certain wavelength $\lambda=2\pi / k$, it can be seen as a single-mode analysis of actual cosmological structure formation.

The original pancake problem has an analytical solution \citep{Anninos94}, which describes the collapse of a pressureless fluid up to caustic formation, happening at redshift $z_{c}$ (the moment of the first shell crossing). For baryonic collapse, it is valid as long as gas pressure is still negligible and can be used to set up the initial conditions. The Lagrangian positions and velocities are given by

\begin{align}
\label{pancake_ic}
\rho(x_l)&=\rho_0\left[1-\frac{1+z_c}{1+z}\cos (kx_l)\right]^{-1}\\ \notag
v(x_l)&=-H_0 \frac{1+z_c}{\sqrt{1+z}}\frac{\sin(kx_l)}{k}
\end{align}

To set up the test, we transform them to Eulerian coordinates:
\begin{align}
x=x_l -\frac{1+z_c}{1+z} \frac{\sin (kx_l)}{k}
\end{align}

This `single pancake' test has been used by many authors to test cosmological hydrocodes, e.g. \citet{Ryu93}, \citet{Gheller96}, \citet{Ricker00}. We skip it here (it will appear again in section \ref{mhdpancake}) and directly move on to the `double pancake' test. It has been proposed by \citet{Bryan95} and not been considered by any other group ever since. In this test, a second wave with one fourth of the wavelength is superimposed on the original wave, utilizing the same formula (\ref{pancake_ic}). The parameters of the two waves are
\begin{align*}
\lambda_1&=64 \;\textrm{Mpc}/h & \lambda_2&=16\;\textrm{Mpc}/h\\
z_{c1}&=1 & z_{c2}&=1.45
\end{align*}
and are evolved from $z=30$ to $z=0$. The initial baryon temperature is set to $T_{init}=13\;\textrm{K}$ according to the formula given in \citet{Anninos96}.

The double pancake is a much more challenging test than the single pancake, not only because it introduces stronger shocks, but also because the superimposed additional wave leads to much finer features which are harder to resolve by the code, especially the temperature peaks. The ratio of thermal to kinetic energy density $\rho\varepsilon/\frac{1}{2}\rho v^2$ in this test covers an extremely wide range between $10^{-9}$ and $10^5$, making it a very stringent test on the correct implementation of the dual energy formalism.

The numerical results are shown in Figure \ref{fig:double_pancake}. While the low-resolution run with $N=64$ fails to recover all features of the solution (the structure of the central density peak, the peak separation in the temperature), they are present in higher resolutions. The high-resolution run with $N=1024$ impressively recovers the solution of \citet{Bryan95}, and due to the dual energy method, the sharp side peaks in the temperature are resolved extremely well. Also, in the extremely cold regions outside of the peaks the temperature is tracked correctly without any oscillations or other artefacts. We could not reproduce this result without our dual energy implementation or with other cosmological codes publically available. We acknowledge that the gravitational solver and the cosmological expansion (through supercomoving coordinates) are implemented correctly and that our variation of the dual energy formalism effectively improves tracking of the temperature without having a dynamical effect on the density or velocity of the gas.

We took a closer look at how well the solution of this test converges. For this, we ran the exact same test as described above, with different resolutions from $N=8$ to $N=512$. As long as the behaviour is linear, that is, well before caustic formation, we can define the relative $\Delta_1$ error norm of a quantity $q$ as:
\begin{align}
\label{Delta_1}
\Delta_1q=\frac{1}{N}\sum_i \frac{|q_i-q_{ref}|}{|q_{ref}|}
\end{align}
The left side of figure \ref{fig:pancake_error} shows this error for the density and velocity as a function of resolution at $z=20$. We took the analytical solution (see above) as reference. Before the calculation of the error, the velocity was shifted by a constant, so that it does not approach zero. We find a constant convergence rate around $N^{-1.1}$ for the whole resolution range.

For the final solution at $z=0$, we are far in the non-linear regime, and the solution features strong discontinuities and even singularities in the density. If we are to make a similar study here, we have to redefine what we take as the error. The analytical solution is not valid at this point, so we compare against a high-resolution run ($N=1024$), which is binned down accordingly, and use the relative self-convergence error:
\begin{align}
\Delta q=\frac{1}{N} \sum_i \frac{|q_i-q_{1024}|}{\max(|q_i|,|q_{1024}|)}
\end{align}
The denominator is chosen this way to force all terms to be between 0 and 1. It leads to more meaningful results, because the differences can span over several orders of magnitude due to the strong discontinuities present. This error is plotted again for density and velocity at $z=0$ (the right side of figure \ref{fig:pancake_error}). Because of the non-linearity of the system, the solution converges much slower at first and reaches $N^{-1}$ only at high resolutions. There is a minimum of resolution required to get the shape of the solution right (around $N=128$), for lower resolutions there are features missing. This is especially the case for the velocity distribution with its pronounced minima and maxima, producing the kink in the convergence rate between $N=64$ and $N=128$, and after that the convergence improves.

When comparing this performance, it turns out that even in \citet{Bryan95}, where the same test is run with a third-order accurate piece-wise parabolic (PPM) code (while our scheme is second-order piece-wise linear), the density distribution converges as $N^{-1.5}$ in the linear regime; and for $z=0$, it does not get better than $N^{-1}$ either. The PPM code of \citet{Ricker00} reaches only $N^{-0.6}$ at $z=7$ for the single pancake. In this context, we can safely state that our code performs adequately well. Being a second-order scheme, the MHD solver requires less computational steps and is faster than higher order methods, while attaining comparable accuracy in the nonlinear regime of structure formation.


\section{Cosmological MHD simulations}
\label{sec:cosmomhd}

Having tested the functionality of the \amiga\ code, we now move on and combine the $N$-body solver, MHD and gravity to perform full cosmological MHD simulations. The aim of this section is to quantify the impact that the introduction of initial large-scale magnetic fields has on simulations of the evolution of dark matter and baryons in a cosmological context.

\subsection{MHD pancake}
\label{mhdpancake}

Before running simulations with realistic cosmological initial conditions, we use the Zel'dovich pancake collapse model in the sense of a single-mode analysis to get an idea of what effects to expect from the presence of a magnetic field. We use a wave in $x$-direction with $z_c=1$ and $\lambda=64\textrm{ Mpc}/h$, but with a few modifications. First, we also want to study the dark matter component. So we include dark matter particles and baryons simultaneously in the simulation, using a baryon fraction $f_b=0.165$ (like later in the full simulations). Both follow the same density and velocity distribution initially. Then, we apply a perpendicular, constant initial magnetic field $\boldsymbol B_{init}$ pointing in $y$-direction. This whole system is then evolved in a flat EdS universe ($\Omega_0=1,\;\Omega_\Lambda=0$) until $z=0$. We repeat the same simulation for a wide range of different initial magnetic fields from $B_{init}=10^{-11}$ G, where the magnetic terms are neglectably small and dynamically unimportant, up to $B_{init}=2\cdot 10^{-6}$ G, where the magnetic field accounts for several percent of the total energy density of the gas.

Figure \ref{fig:MHDpancake} shows the density and temperature of the baryons, the distribution of the dark matter particles and the magnetic field strength at $z=0$ for different runs. Initial fields up to up to 0.05 $\mu$G do not have any significant effect on the density profiles (and the other quantities) and the field strength just follows the density profile of the baryons. Higher fields, however, induce large changes at the shock and post-shock regions, slowing down the baryon collapse and smearing out the baryon density profile. High field strengths effectively prevent the build-up of sharp, high-temperature baryon peaks. Although the situation is not directly transferrable to a full 3D simulation, this study gives us a hint on the general behaviour of density peak formation under the influence of a magnetic field.

The dark matter distribution is generally much less affected than the baryon component -- it does not interact directly with the magnetic field, but only indirectly through the gravitational force of the baryons. Since the dark matter particles are collisionless, they do not clump all together in the centre, but pass through each other (this happens exactly at $z=z_c$) and form side peaks. The shape of the central dark matter peak gets somewhat distorted if the baryons clump differently due to the field, but the height and position of the side peaks that form after $z_c$ lie outside the shocked regions and are practically not affected.

\begin{figure}
\begin{center}
\begin{minipage}{0.485\textwidth}
        \epsfig{file=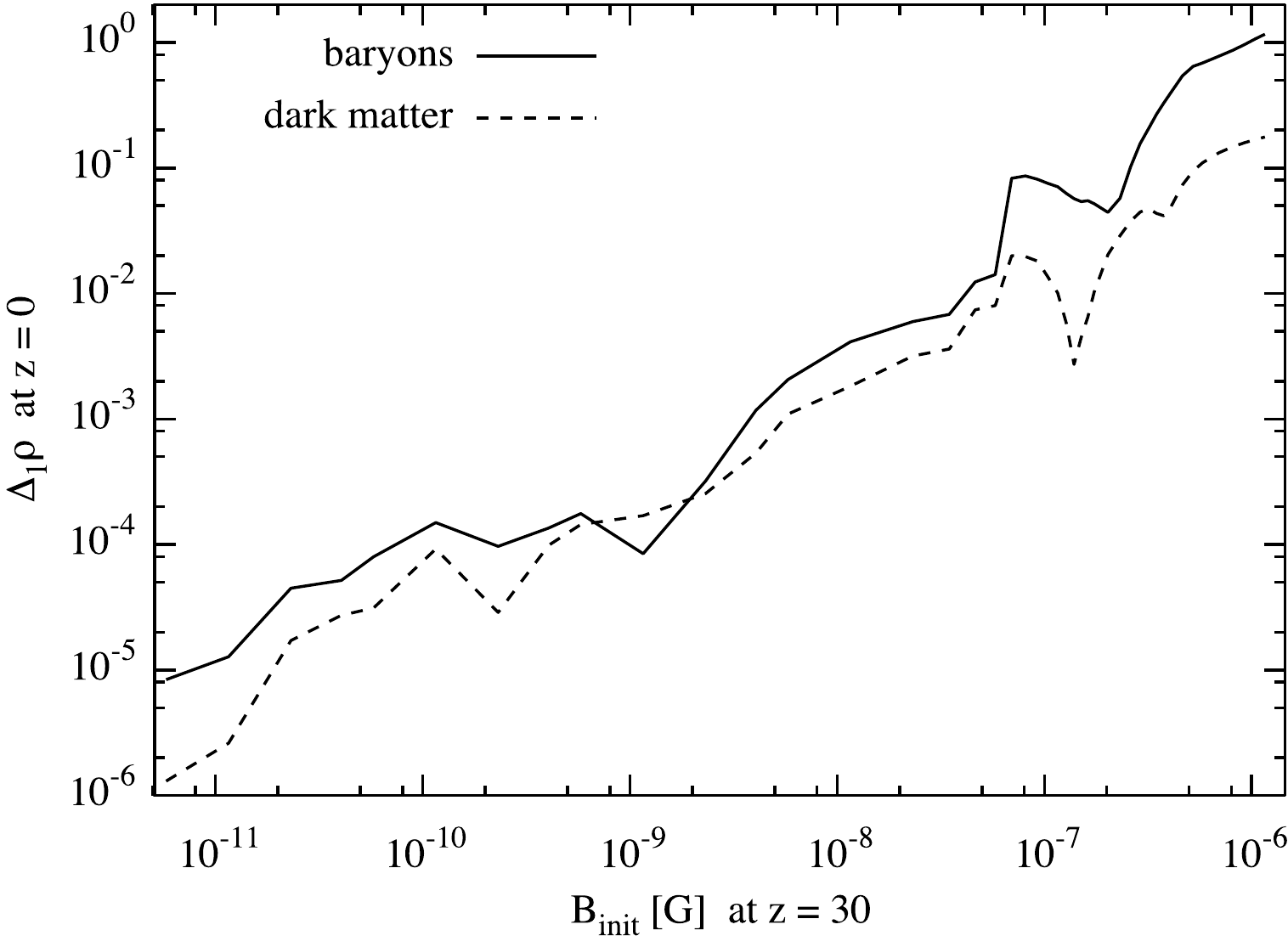, width=1\textwidth, angle=0}
\end{minipage}
\end{center}
\caption{Relative deviation ($\Delta_1$ norm) of the final baryon and dark matter density distribution from the $B_{init}=0$ case depending on $B_{init}$ in the numerical MHD pancake solution.
\label{fig:MHDpancake2}}
\end{figure}

For a more quantitative view on these effects, we took the non-magnetic numerical pancake solution at $z=0$ as a reference and calculated the deviation created by an initial magnetic field as the relative $\Delta_1$ ``error'' norm (\ref{Delta_1}) of baryon and DM density distributions over the interval of interest $x \in [0.43,0.57]$, a quantity very sensitive to numerical deviations (Figure \ref{fig:MHDpancake2}). Expectedly, the deviation rises with higher initial fields, and the dark matter is less affected than the baryons. It should be noted that there is a certain level of numerical noise: If one changes a code parameter, like the CFL number, initial timestep, or dual energy parameters (within reasonable values of course), the relative deviation will be typically around $\Delta_1 \rho \approx 10^{-4}$. Smaller deviations can therefore be considered statistically insignificant. We can see that, in order to have a significant effect of, say, $\Delta_1 \rho = 1\%$, the energy density of the initial magnetic field at $z=30$ has to be at least around $10^{-7}$G.


\begin{figure*}
\begin{center}
\begin{minipage}{0.9\textwidth}
        \epsfig{file=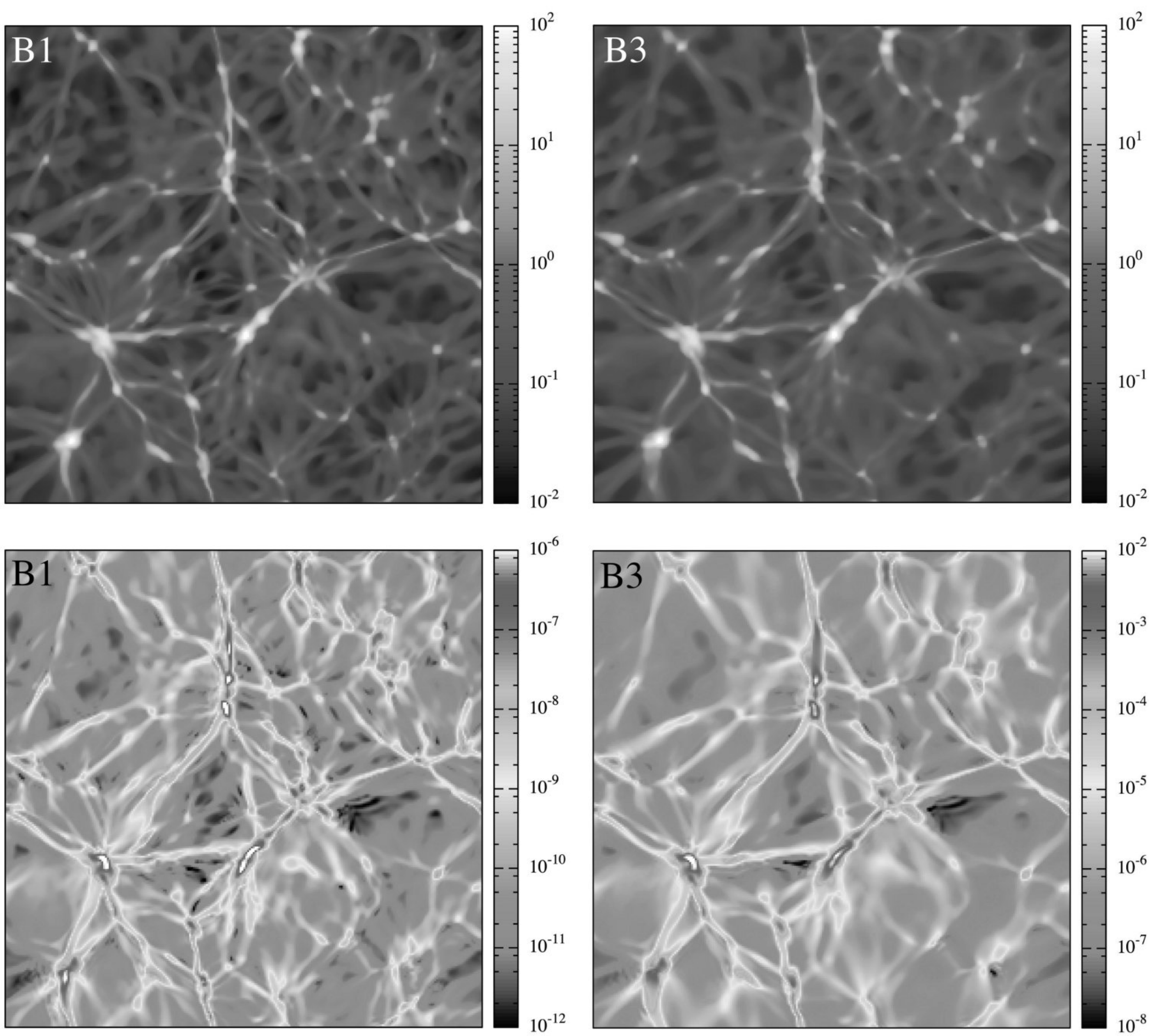, width=1\textwidth, angle=0}
\end{minipage}
\end{center}
\caption{Comparison of cosmological MHD simulations at $z=0$ with low and high initial magnetic fields. Shown is the baryon density (top) and magnetic energy density (bottom) in code units. The maps are cuts through the simulation box in the $x,y$ plane at half box depth.
\label{fig:MHD256maps}}
\end{figure*}


\subsection{Cosmological MHD simulations}
\label{cosmoMHDsimulations}

\subsubsection{The initial magnetic field}

In order to conduct simulations of the cosmic structure formation that take into account the primordial magnetic field, one must choose appropriate initial conditions. Yet, the possible strength\footnote{When talking about primordial field strengths at earlier times (higher redshifts), we mean the \emph{comoving} magnetic field strength $B_{comoving}=a^{-2}B_{proper}$, that is, the strength such a field would have when extrapolated to the present-day scale factor. It is convenient because it allows for the magnetic field strengths from different epochs to be compared directly. The relation $B(z)=B_0/a^2$ follows from magnetic flux conservation.}, shape or origin of a primordial field remains unclear at the moment. Theories on this subject suggest that a cosmological large-scale magnetic field was already present before recombination ($z\sim 1000$). Such a primordial magnetic field could have been produced during inflation \citep{Turner88,Gasperini06}, much like the primordial density fluctuations that led to structure formation, or during subsequent phase transitions \citep{Gopal05}. Unfortunately, the predictions of such models involving string theory and particle physics are presently highly parameter dependent and rather inconclusive. Certain models can lead to fields as large as 1 nG, while others predict fields that are many orders of magnitude smaller \citep{Subramanian08}.

While at the present there is no working theory to estimate the possible strength of a primordial magnetic field, it is feasible to give some constraints. If a magnetic field coherent at cosmological scales was present in the early universe, it should have left its imprint on the linear polarization of the CMB by Faraday rotation. Based on that, \citet{Kahniashvili09} derived an upper limit for a primordial magnetic field based on the CMB polarization power spectrum from the WMAP 5-year data. They find that at a scale of 100 Mpc, the field amplitude must have been smaller than 0.7 nG. At smaller scales of 1 Mpc, the upper limit may be as high as 30 nG, depending on the assumed power spectrum.

Another way to constrain the primordial magnetic field stems from Big Bang Nucleosynthesis (BBN). The presence of a magnetic field during BBN would have changed the nuclear reaction rates, thus resulting in an altered abundance of lighter elements like ${}^{3}$He, ${}^{4}$He, ${}^{7}$Li. To be compatible with the current agreement between BBN theory and element abundancy observations, a primordial field must be smaller than some critical value. First constraints derived in that manner were pretty high, up to 1 $\mu$G \citep{Kernan96}; later, \citet{Grasso01} deduced a more realistic value of 1 nG for Mpc-scale fields with the help of some additional assumptions.

For the simulations presented in the following section, we assume a primordial field already present before the starting time of the simulation. We further assume it to be constant and homogeneous in the whole simulation box, since the focus lies on how structure formation is affected by magnetic fields on scales larger than individual structures. It could be argued that a homogeneous primordial field pointing in one direction contradicts the assumption of an isotropic universe by creating a direction of preference; but the anisotropy created by such initial conditions has no impact on a statistical analysis of the baryon evolution, because the angle between the field vector and the baryon flow, a crucial quantity for the magnetic force on the baryons, is still randomly distributed, and the magnetic pressure does not depend on the field direction at all. It is also worth noticing that ideal MHD predicts the magnetic field lines to follow the baryon distribution. In fact, gravitationally collapsing baryonic structures would completely reshape the magnetic field distribution up to the point that any information on the original shape of the primordial field would be lost \citep{Dolag02}, so the initial field shape should not significantly influence the results.

\subsubsection{Overview and initial conditions}

For this study, we carried out a set of cosmological 3D simulations with varying primordial field strengths $B_{init}$. The simulations model a universe containing baryons and dark matter particles in a three-dimensional 64 Mpc$/h$ box with periodic boundary conditions, using a baryon fraction of $f_b=0.165$. The initial conditions used for all of the simulation runs were created from an initial CDM power spectrum corresponding to the WMAP 5 parameters \citep{Komatsu09}: $\Omega_0=0.273$ and $\Omega_\Lambda=0.726$ with the \textsc{PMCODE} IC package \citep{Klypin97}. The initial density distribution of the baryons follows the dark matter. The initial field is set to a constant magnetic field in $y$-direction: $\boldsymbol B_{init} = (0,B_{init},0$). Apart from the different initial magnetic field strengths, the initial conditions are identical for all the runs. The simulations were run from the chosen starting redshift $z_{init}=30$ until $z=0$ with the full MHD version of the \amiga\ code on a regular $256^3$ grid utilizing OpenMP statements for parallelization.

\begin{table}
\begin{center}
\begin{minipage}{0.485\textwidth}
\begin{tabular}{|l|l|l|}
\hline
Simulation & Initial comoving magnetic & Initial \emph{physical} magnetic  \\
& field strength $B_{init}$ at $z=30$ & field strength at $z=30$ \\
\hline
B0 & 0 G & 0 G \\
B1 & 5.79 $\cdot 10^{-10}$ G & 5.56 $\cdot 10^{-7}$ G \\
B2&5.79 $\cdot 10^{-9}$ G& 5.56 $\cdot 10^{-6}$ G \\
B3&5.79 $\cdot 10^{-8}$ G& 5.56 $\cdot 10^{-5}$ G \\
\hline
\end{tabular}
\end{minipage}
\end{center}
\caption{Identifiers and initial magnetic field strengths for the MHD simulations used in this paper
\label{tab:overview}}
\end{table}

The initial magnetic field values are summarized in Table 1. While the lowest initial field (B1) is compatible with current theoretical and observational estimates for a large-scale field, the highest initial field (B3) is significantly higher than all current upper limits. The energy density of the B1 field at $z=30$ is equivalent to the kinetic energy density of a gas with the average gas density at that redshift moving at 0.4 km/s. This is clearly too small to be dynamically important at any stage of the simulation. The B3 field, on the other hand, is $10^2$ times stronger and has $10^4$ times more energy density, so we should expect an effect due to its presence.


\begin{figure}
\begin{center}
\begin{minipage}{0.485\textwidth}
        \epsfig{file=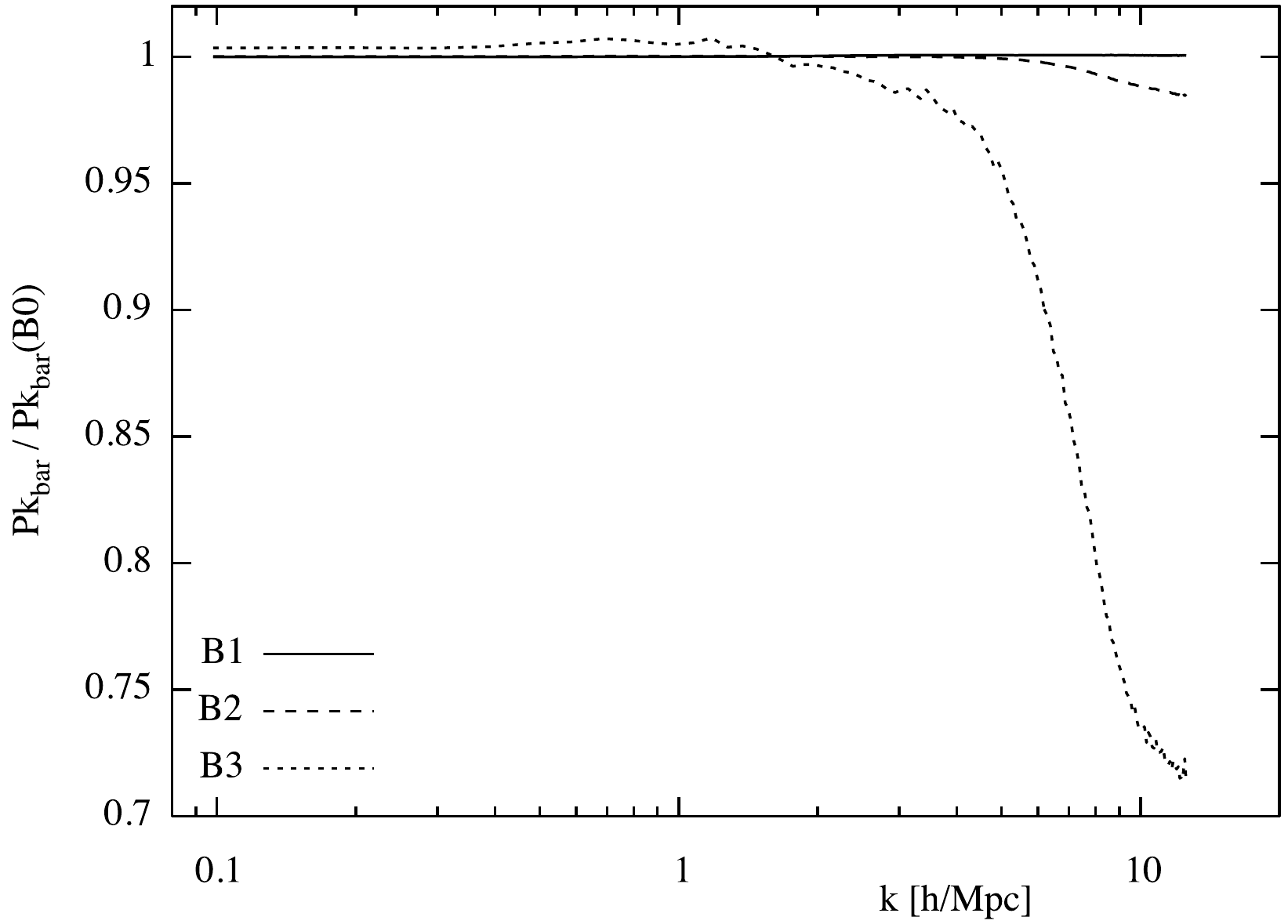, width=1\textwidth, angle=0}
\end{minipage}
\end{center}
\caption{Power spectrum of baryons at $z=0$ for MHD runs with different initial magnetic fields relative to the power spectrum without magnetic field. For sufficient field strengths, the power spectrum of the baryon distribution shows a suppression of finer structures (the dark matter power spectrum stays unchanged).
\label{fig:relativePk}}
\end{figure}

\begin{figure}
\begin{center}
\begin{minipage}{0.485\textwidth}
        \epsfig{file=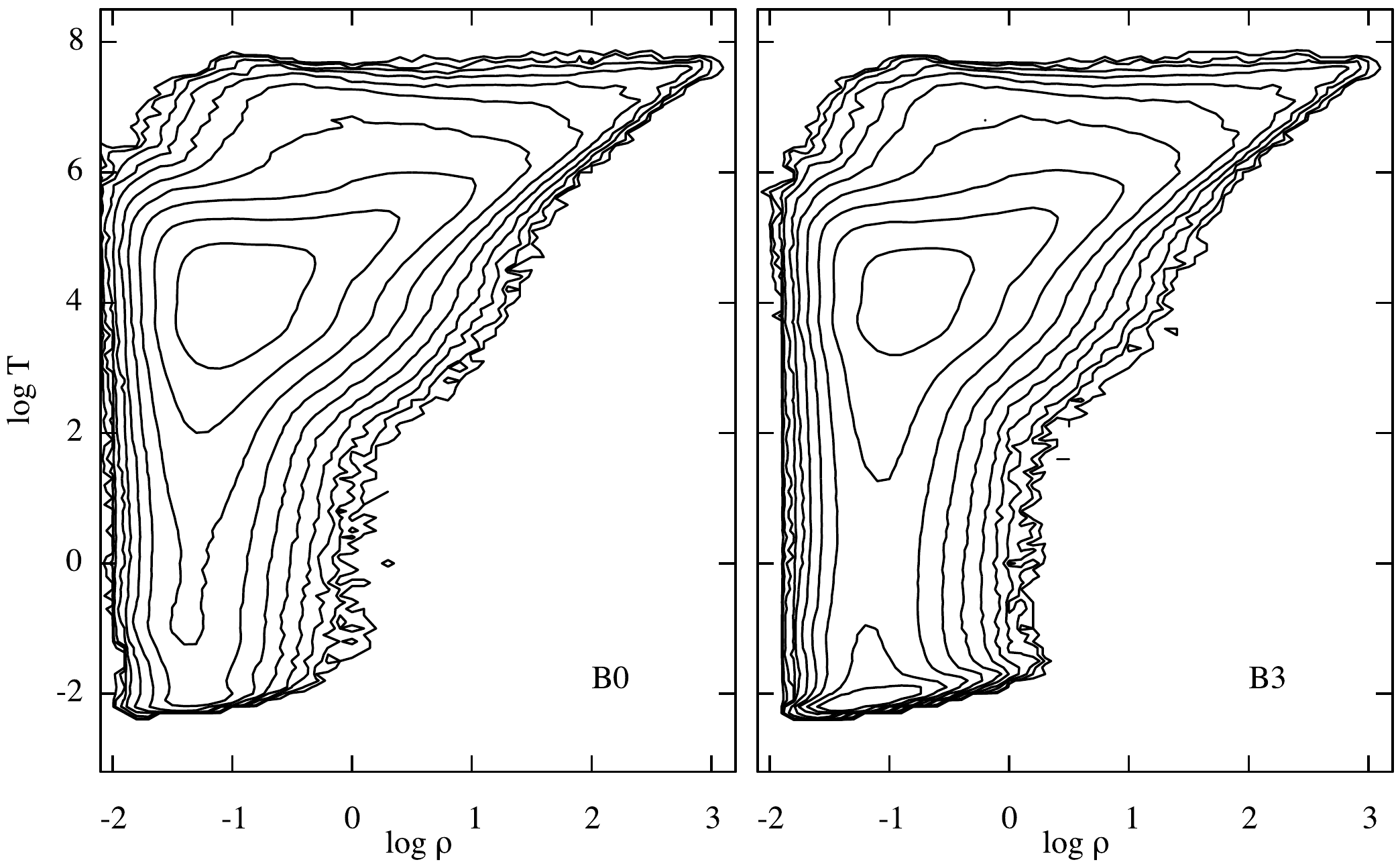, width=1\textwidth, angle=0}
\end{minipage}
\end{center}
\caption{Contour plot of the volume fraction with given baryon temperature and density at $z=0$ without and with a strong initial magnetic field.
\label{fig:contourplot}}
\end{figure}


\subsubsection{Magnetic field influence on the large-scale distribution of baryons}

Figure \ref{fig:MHD256maps} shows maps of the baryon density and magnetic energy density at the final redshift $z=0$ of our simulations, with the weakest and strongest initial magnetic field that was simulated. Even for the very high magnetic field of the B3 run, the baryon distribution does not change significantly; a closer look reveals that the distribution becomes slightly smeared out, featuring less fine structure. The weaker initial fields do not have an influence at all. The magnetic field shows the expected behaviour: in ideal MHD, it is frozen into the gas motion and largely follows the gas distribution.

In the following, we want to analyse quantitatively whether the presence of magnetic fields in these simulations alters the result for the other constituents.

The smearing of structural features by magnetic fields can be quantified by looking at the power spectrum at $z=0$. In figure \ref{fig:relativePk} we plot the power spectrum $P_{bar}(k)$ of the baryons relative to the power spectrum of the non-magnetic run. It can be seen that the additional magnetic pressure and the altering of baryon flows by magnetic tension leads to a characteristic suppression of structure formation at finer scales. This is, however, a very small effect that becomes important only at smaller scales and high magnetic fields. The high initial field in the B3 run lowers the power spectrum by 10 \% on a scale of 1 Mpc$/h$, the weaker field in the B2 model only has a (very slight) effect below 1 Mpc$/h$, and the more realistic field of the B1 model has no effect whatsoever. On scales above a few Mpc$/h$, no influence can be seen even with the strongest field in our simulation runs. On scales well below 1 Mpc$/h$, additional processes inside the individual collapsing structures become important, especially in their core regions. There, the magnetic field is influenced by cooling flows, turbulence and field tangling, which is not resolved by the large-scale runs conducted here. They can be better addressed by finer simulations of individual objects like the study of \citet{Dubois08} on a single magnetized galaxy cluster.

Figure \ref{fig:contourplot} shows the number of cells with a certain value of baryon density $\rho$ and temperature $T$ within the simulated box, again for the strongest initial field and without a field. While the latter shows the characteristic shape known from other cosmological codes (see e.g. \citealt{Ryu93} for similar figures), a strong magnetic field leads to an additional maximum at the bottom, where regions with very cool gas are located. We can compare this directly to the result for a single Zel'dovich wave in Figure \ref{fig:MHDpancake}, where a similar effect occurs: a broadening and smearing of the density profile and the formation of a cool region behind the shock front. These results are in good agreement with \citet{Gazzola07}, where the same smoothing of the mass distribution with shallower density profiles and ``washed out'' finer density clumps can be seen, although they add the magnetic field simply as an additional isotropic pressure term instead of a proper MHD treatment. In any case, for magnetic fields of order $\sim1$ nG and below, no effect whatsoever can be seen on the scales resolved by our simulation.

To summarize, on scales of $\sim1$ Mpc and above, only primordial fields significantly higher than of order $\sim$ nG have a noticeable effect on the baryon dynamics and gas distribution, which can be safely stated to be outside the upper theoretical and observational limits.

\section{Summary and Conclusions}
\label{sec:summary}

In this paper, we present the new numerical $C$ code \amiga\, designed to perform cosmological magnetohydrodynamic simulations. It contains the powerful and memory-efficient AMR $N$-body code from its predecessor \mlapm\ \citet{Knebe01}, as well as a newly developed Eulerian grid-based MHD solver based on \citet{Ziegler04} and \citet{Ziegler05}. The new code allows to simulate dark matter, baryon physics and magnetic fields in a self-consistent way inside a full cosmological framework. To facilitate the numerical solution of cosmological MHD equations, the code is working with \emph{supercomoving} coordinates, a transformation that greatly simplifies the equations, while preserving the fully cosmological setting. There are implemented techniques to properly resolve strong shockwaves and supersonic flows in the baryon component, and to ensure the important condition of a divergence-free magnetic field down to machine precision. By conducting a series of test problems we acknowledge the high accuracy of this new code.

As a first application of the new code, we present simulations of the cosmic structure formation with primordial magnetic fields. Such large-scale magnetic fields, possibly of cosmological origin, can be expected from different theoretical models and observational evidence of magnetic fields inside galaxy clusters. We want to address the question whether they could be a relevant factor for the large-scale dynamics in cosmological simulations.

The simulations carried out with \amiga\ model a $\Lambda$CDM universe with the WMAP-5 cosmology in a comoving 64 Mpc$/h$ computational volume, and its evolution from redshift $z_{init}=30$ to $z=0$. The applied primordial field strengths range from about 0.5 nG (a likely value from current constraints) to about 50 nG, which is significantly higher than current theoretical and observational constraints for magnetic fields on such large scales. The analysis of the simulations reveals that only in this last case, a large-scale magnetic field has a statistically significant influence on the baryon dynamics. Then, the magnetic pressure and tension leads to a suppression of baryonic small-scale structure and smears out density peaks, visible in the baryonic power spectrum. However, even the highest simulated initial field has no noticeable effect on scales above a few Mpc$/h$. We can therefore conclude that, since current theoretical and observational constraints predict a large-scale field not much stronger than $\sim$ 1 nG, at least outside of the core regions of gravitationally collapsed structures it cannot have a significance for the baryonic component during large-scale structure formation, neither on the power spectrum nor on the actual distribution.

Even though our simulations do not have the required resolution to study the (internal) properties of individual objects, we nevertheless like to close with a brief discussion of our findings in that direction. We observed (though not explicitly presented here) that the mass function of collapsed structures remains unaffected even for magnetic fields as large as the ones in model B3. Furthermore, the shape of (dark matter) haloes also appeared unaltered when increasing the strength of the primordial magnetic field. And for the baryon fraction -- for which \citet{Gazzola07} have shown a dependence on the magnetic field strength -- our own results are unfortunately affected by resolution effects: while stronger magnetic fields lead to a depletion of baryons in smaller mass objects (cf. Figure 8 in \citet{Gazzola07}), the same is caused by a lack of resolution in cosmological codes (\citealt{Crain07,Rudd08}); hence, we observe this effect but attribute it to our resolution. Further studies and more refined simulations in this direction are necessary to clarify this subject in greater detail.


\section*{Acknowledgements}
TD acknowledges support through the AstroSim network of the European Science Foundation (Exchange Grant 2496). AK acknowledges funding through the Emmy Noether programme of the DFG (KN 755/1). AK is further supported by the Ministerio de Ciencia e Innovaci\'on (MICINN) in Spain through the Ramon y Cajal programme. We further acknowledge the LEA Astro-PF collaboration and the AstroSim network (Science Meeting 2387) for the financial support of the workshop "The local universe: from dwarf galaxies to galaxy clusters" held in Jab\l onna near Warsaw in June/July 2009, during which a part of this work has been conducted. The simulations presented herein have been performed on the \textsc{Babel} cluster at the Astrophysical Institute Potsdam (AIP). The authors wish to thank Jochen Klar from AIP for providing the analytical reference solutions for the hydrodynamic test cases. 

\bibliography{archive} \bsp

\label{lastpage}

\newpage

\begin{onecolumn}

\renewcommand{\theequation}{A\arabic{equation}}
  \setcounter{equation}{0}  

\section*{Appendix: Derivation of the supercomoving MHD equations}

As the starting point for the supercomoving transformation we take the full set of equations describing dark matter, baryons and magnetic fields in the ordinary non-comoving frame:
\begin{subequations}
\begin{align}
\label{proper_a}
&\frac{\textrm{d} \boldsymbol r_{dm}}{\textrm{d}t}=\boldsymbol v_{dm} \\
\label{proper_b}
&\frac{\textrm{d} \boldsymbol v_{dm}}{\textrm{d}t}=-\boldsymbol\nabla\phi\\
\label{proper_c}
&\Delta\phi=4\pi G\rho_{tot}\\
\label{proper_d}
&\frac{\partial \rho}{\partial t}+\boldsymbol\nabla\cdot(\rho \boldsymbol v)=0 \\
\label{proper_e}
&\frac{\partial (\rho \boldsymbol v)}{\partial t}+\boldsymbol\nabla\cdot\left [\rho \boldsymbol v \boldsymbol v + \left(p+\frac{B^2}{2\mu}\right)I-\frac{1}{\mu} \boldsymbol B \boldsymbol B\right]=-\rho\, \boldsymbol\nabla\phi\\
\label{proper_f}
&\frac{\partial (\rho E)}{\partial t}+\boldsymbol\nabla\cdot\left[\boldsymbol v \left(\rho E+p+\frac{B^2}{2\mu}\right)-\frac{1}{\mu}\boldsymbol B(\boldsymbol v\cdot\boldsymbol B)\right]=-\rho \boldsymbol v \cdot(\boldsymbol\nabla\phi)\\
\label{proper_g}
&\frac{\partial \boldsymbol B}{\partial t}+\boldsymbol\nabla\times(-\boldsymbol v \times\boldsymbol B)=0\\
\label{proper_h}
&\boldsymbol\nabla\cdot\boldsymbol B=0 \\
\label{proper_i}
&\frac{\partial S}{\partial t}+\boldsymbol\nabla\cdot(S \boldsymbol v)=0
\end{align}
\end{subequations}
Here, equations (\ref{proper_a}) and (\ref{proper_b}) are the equations of motion for the dark matter ($dm$) particles; equation (\ref{proper_c}) is Poisson's equation, where $\rho_{tot}$ is the total (dm + baryons) density; and the equations (\ref{proper_d}) -- (\ref{proper_h}) are the equations of ideal MHD. For the dual energy formalism, we also need to transform the equation (\ref{proper_i}) describing the modified entropy. Below, we will apply the supercomoving transformation to each of these equations individually and construct a new set of supercomoving equations, using the definitions (\ref{definition}) and (\ref{transformation}). The new supercomoving quantities and derivatives will be denoted by a subscript $x$. Throughout this appendix, an overdot denotes the temporal derivative with respect to the proper, non-supercomoving time $t$.


\subsection{Dark matter particle equations of motion}

We define the supercomoving velocity $\boldsymbol v_{x,dm}$ as the derivative of $\boldsymbol x_{dm}$ with respect to the supercomoving time $\textrm{d}t_x=\textrm{d}t/a^2$:
\begin{align}
\frac{\textrm{d}\boldsymbol x_{dm}}{\textrm{d}t_x}=\boldsymbol v_{x,dm}
\end{align}
From this follows the relation between physical and supercomoving velocity:
\begin{align}
\boldsymbol v_{x,dm} = a \boldsymbol v_{dm} - \dot a \boldsymbol r_{dm} \;\;\;\;\;\Longleftrightarrow \;\;\;\;\; \boldsymbol v_{dm} = \frac{1}{a}\boldsymbol v_{x,dm} + \dot a \boldsymbol x_{dm}
\end{align}
It follows also from the definition that the spatial derivatives change to
\begin{align*}
\boldsymbol\nabla_x=a\boldsymbol\nabla\;\;\;;\;\;\;\Delta_x = a^2 \Delta\;\;\;.
\end{align*}
The goal is to obtain an equation of motion analogous to equation \ref{proper_b} for the supercomoving dark matter velocities $\boldsymbol v_{x,dm}\;$. Let us consider the supercomoving acceleration:
\begin{align*}
\frac{\textrm{d}\boldsymbol v_{dm}}{\textrm{d}t_x}=\frac{\textrm{d}}{\textrm{d}t_x}\left( \frac{\textrm{d}\boldsymbol x_{dm}}{\textrm{d}t_x} \right)
\end{align*}
Now we replace the supercomoving time and position with the physical time and position, and rewrite the result to obtain a relation with the supercomoving gravitational force:
\begin{align*}
\frac{\textrm{d}\boldsymbol v_{x,dm}}{\textrm{d}t_x}&=a^2\frac{\textrm{d}}{\textrm{d}t}(\boldsymbol v_{x,dm}) \\ 
&=a^2\frac{\textrm{d}}{\textrm{d}t}(a \boldsymbol v_{dm} - \dot a \boldsymbol r_{dm})\displaybreak[0] \\
&=a^2 \left(a\frac{\textrm{d}\boldsymbol v_{dm}}{\textrm{d}t}+\dot a \boldsymbol v_{dm} -\ddot a \boldsymbol r_{dm}- \dot a \boldsymbol v_{dm}\right )\displaybreak[0]\\
&=a^2 \left(a\frac{\textrm{d}\boldsymbol v_{dm}}{\textrm{d}t}-\ddot a \boldsymbol r_{dm} \right )\displaybreak[0]\\
&=a^2\left(a\frac{\textrm{d}\boldsymbol v_{dm}}{\textrm{d}t}-a\ddot a \boldsymbol x_{dm}\right)\displaybreak[0] \\
&=a^2\left[ -\boldsymbol \nabla_x \phi + a \ddot a \left (-\boldsymbol \nabla_x \left(\frac{1}{2} x_{dm}^2\right)\right) \right] \displaybreak[0] \\
&=-\boldsymbol\nabla_x\left[ a^2 \left( \phi + \frac{1}{2}a \ddot a x^2_{dm}\right) \right] \\
&=-\boldsymbol\nabla_x \phi_x
\end{align*}
Here, we used equation (\ref{proper_b}) and the definition of the comoving potential $\phi_x\,$. Now we see that the supercomoving equations of motion are formally identical to their proper physical counterparts, although the quantities are defined differently:
\begin{subequations}
\begin{align}
\frac{\textrm{d}\boldsymbol x_{dm}}{\textrm{d}t_x}&=\boldsymbol v_{x,dm} \\
\frac{\textrm{d}\boldsymbol v_{x,dm}}{\textrm{d}t_x}&=-\boldsymbol\nabla_x \phi_x
\end{align}
\end{subequations}
This is the main advantage of supercomoving coordinates over the comoving coordinates, which explicitly include additional factors depending on $a$. We will see that the other equations behave in a similar way.


\subsection{Poisson's equation}

Poisson's equation determines the potential $\phi$ of the system, and as such it is the only equation where cosmology enters explicitly. Although equation (\ref{proper_c}) describes the gravitational potential in an ordinary physical setting, in a cosmological framework we also have to consider the cosmological constant $\Lambda$. This is realized by adding a $\Lambda$ term that has the dimension of a density, and then using this ``effective density'' in Poisson's equation.

Let us consider the second Friedmann equation, which relates the average total mass density $\bar\rho_{tot}$ and the cosmological constant to the accerelation of the cosmic expansion:
\begin{align}
\frac{\ddot a}{a}= -\frac{4 \pi G}{3}\left( \bar \rho_{tot} + \frac{3p}{c^2} \right) + \frac{\Lambda c^2}{3}
\end{align}
Dark matter is by definition pressureless $(p=0)$, and the pressure of the small baryonic component can be neglected. Then we can write
\begin{align}
\label{effectivefriedmann2}
\frac{\ddot a}{a}= -\frac{4 \pi G}{3}\left( \bar \rho_{tot} - \rho_\Lambda \right )
\end{align}
with $\rho_\Lambda=-\Lambda c^2/4\pi G$. Then, if the cosmological constant is not zero, Poisson's equation effectively becomes
\begin{align}
\label{poissonwithlambda}
\Delta\phi=4\pi G (\rho_{tot}-\rho_\Lambda)\;\;\;.
\end{align}
We formulate the left-hand side in terms of the supercomoving gravitational potential:
\begin{align*}
\Delta \phi &= \frac{1}{a^2}\Delta_x\left( \frac{\phi_x}{a^2} - \frac{1}{2} a \ddot a x^2 \right)\\
&= \frac{1}{a^4}\Delta_x\phi_x -\frac{\ddot a}{2 a}  \Delta_x x^2\\
&= \frac{1}{a^4}\Delta_x\phi_x -3\frac{\ddot a}{a} 
\end{align*}
Using again the second Friedmann equation \ref{effectivefriedmann2}, we get
\begin{align*}
\Delta \phi &= \frac{1}{a^4}\Delta_x\phi_x -3\left[\frac{-4\pi G}{3} (\bar\rho_{tot}-\rho_\Lambda) \right] \;\;\;.
\end{align*}
Equating this with the right-hand side of equation \ref{poissonwithlambda} yields
\begin{align*}
\frac{1}{a^4}\Delta_x\phi_x -4\pi G (\bar\rho_{tot}-\rho_\Lambda)&=-4\pi G(\rho_{tot}-\rho_\Lambda) \;\;\;.
\end{align*}
The $\Lambda$ term cancels, and we arrive at the supercomoving Poisson's equation:
\begin{align}
\notag
\Delta_x \phi_x &= 4\pi G a^4 (\bar\rho_{tot}-\rho_{tot})\\
&= 4\pi G a (\bar\rho_{x,tot}-\rho_{x,tot})\;\;\;.
\end{align}
The supercomoving version of Poisson's equation looks slightly different than the non-cosmological one: the density contrast enters instead of the total density, because the supercomoving potential is responsible for peculiar motions due to density fluctuations, while the total density governs the overall expansion.


\subsection{Baryon mass density}
Now we will transform the conservation law for the baryon density $\rho$,
\begin{align*}
\frac{\partial \rho}{\partial t}+\boldsymbol \nabla (\rho \boldsymbol v)=0\;\;\;.
\end{align*}
First, we replace proper time, density and velocity with their comoving counterparts. We begin by replacing the partial time derivative $\partial/\partial t$ at constant position $\boldsymbol r$ with the one at constant comoving position $\boldsymbol x$:
\begin{align*}
\frac{\partial}{\partial t}\bigg|_r \rho&=\frac{\partial}{\partial t}\bigg|_x \rho-\frac{\dot a}{a}\boldsymbol x\cdot \boldsymbol \nabla_x \rho\\
&=\frac{1}{a^2}\frac{\partial}{\partial t_x} \left(\frac{1}{a^3}\rho_x\right)-\frac{\dot a}{a^4}\boldsymbol x\cdot\boldsymbol \nabla_x \rho_x \displaybreak[0]\\
&=\frac{-3\dot a}{a^4}\rho_x+\frac{1}{a^5}\frac{\partial\rho_x}{\partial t_x} - \frac{\dot a}{a^4} \boldsymbol x \cdot\boldsymbol \nabla_x\rho_x
\end{align*}
Evaluating the flux term, using $\boldsymbol v=\boldsymbol v_x/a + \dot a \boldsymbol x$:
\begin{align*}
\boldsymbol \nabla (\rho \boldsymbol v) &= \frac{1}{a} \boldsymbol  \nabla_x \left( \frac{1}{a^4} \rho_x \boldsymbol v_x + \frac{\dot a}{a^3}\rho_x \boldsymbol x \right) \\
& = \frac{1}{a^5} \boldsymbol \nabla_x (\rho_x \boldsymbol v_x ) + \frac{\dot a}{a^4} \boldsymbol \nabla_x(\rho_x \boldsymbol x) \displaybreak[0]\\
& = \frac{1}{a^5} \boldsymbol \nabla_x (\rho_x \boldsymbol v_x ) + \frac{\dot a}{a^4} \boldsymbol x (\boldsymbol \nabla_x \rho_x) + \frac{3\dot a}{a^4} \rho_x
\end{align*}
Putting these two expressions together, four of the six terms cancel, leaving only
\begin{align}
\frac{\partial \rho_x}{\partial t_x} + \boldsymbol \nabla_x(\rho_x \boldsymbol v_x)=0\;\;\;.
\end{align}


\subsection{Baryon momentum density}
The conservation law for the baryon momentum can be written as:
\begin{align}
\label{momentumwithA}
\frac{\partial( \rho \boldsymbol v)}{\partial t}+\boldsymbol\nabla\cdot(\rho \boldsymbol v \boldsymbol v +A)=-\rho\, \boldsymbol\nabla\phi\
\end{align}
where we used the abbreviation $A=\left(p+\frac{B^2}{2\mu}\right)I-\frac{1}{\mu} \boldsymbol B \boldsymbol B$. From the definitions of the supercomoving pressure $p_x$ and magnetic field $\boldsymbol B_x$ one immediately sees that
\begin{align*}
A_x = \left(p_x+\frac{B_x^2}{2\mu}\right)I-\frac{1}{\mu} \boldsymbol B_x \boldsymbol B_x= a^5 A\;\;\;.
\end{align*}
We decompose the first term of the momentum equation into two parts:
\begin{align*}
\frac{\partial( \rho \boldsymbol v)}{\partial t}\bigg|_r=\rho \frac{\partial \boldsymbol v}{\partial t}\bigg|_r+\boldsymbol v \frac{\partial \rho}{\partial t}\bigg|_r
\end{align*}
The first part equals
\begin{align*}
\rho \frac{\partial \boldsymbol v}{\partial t}\bigg|_r &= \frac{1}{a^3}\rho_x\left( \rho \frac{\partial \boldsymbol v}{\partial t}\bigg|_x - \frac{\dot a}{a} \boldsymbol x ( \boldsymbol \nabla_x\boldsymbol u) \right)\\
\end{align*}
Using the abbreviation $K=\frac{\dot a}{a} \boldsymbol x ( \boldsymbol \nabla_x\boldsymbol u) $ , it evaluates to
\begin{align*}
\rho \frac{\partial \boldsymbol v}{\partial t}\bigg|_r &= \frac{1}{a^3}\rho_x\left( \rho \frac{\partial \boldsymbol v}{\partial t}\bigg|_x - K\right)\\
&= \frac{1}{a^3}\rho_x\left[ \frac{\partial }{\partial t}\bigg|_x  \left( \frac{1}{a} \boldsymbol v_x + \dot a \boldsymbol x\right) - K\right ] \\
&=\frac{1}{a^3}\rho_x\left( \frac{-\dot a}{a^2} \boldsymbol v_x +\frac{1}{a^3}\frac{\partial \boldsymbol v_x}{\partial t_x} + \ddot a \boldsymbol x - K \right)
\end{align*}
while we write the second part as (equation \ref{proper_b}):
\begin{align*}
\boldsymbol v \frac{\partial \rho}{\partial t}\bigg|_r = - \boldsymbol v \boldsymbol \nabla \cdot (\rho \boldsymbol v)
\end{align*}
The second term on the left-hand side of equation (\ref{momentumwithA}) transforms as follows:
\begin{align*}
\boldsymbol \nabla (\rho \boldsymbol v\boldsymbol v+A) &= \boldsymbol v \boldsymbol \nabla \cdot (\rho \boldsymbol v) + \rho \boldsymbol v \cdot \boldsymbol \nabla \boldsymbol v + \boldsymbol \nabla A \\
&=\boldsymbol v \boldsymbol \nabla \cdot (\rho \boldsymbol v) + \frac{1}{a^3}\rho_x \left[ \left( \frac{1}{a} \boldsymbol v_x + \dot a \boldsymbol x\right ) \cdot \frac{1}{a} \boldsymbol \nabla_x \left( \frac{1}{a} \boldsymbol v_x + \dot a \boldsymbol x\right ) + K \right] + \frac{1}{a^6} \boldsymbol \nabla_x A_x \\
&=\boldsymbol v \boldsymbol \nabla \cdot (\rho \boldsymbol v) + \frac{1}{a^3}\rho_x \left[ \frac{1}{a^3} (\boldsymbol v_x \cdot \boldsymbol \nabla_x) \boldsymbol v_x + \frac{\dot a}{a^2} (\boldsymbol v_x \cdot \boldsymbol \nabla_x)  \boldsymbol x + K \right] + \frac{1}{a^6} \boldsymbol \nabla_x A_x \\
&=\boldsymbol v \boldsymbol \nabla \cdot (\rho \boldsymbol v) + \frac{1}{a^3}\rho_x \left[ \frac{1}{a^3} (\boldsymbol v_x \cdot \boldsymbol \nabla_x) \boldsymbol v_x + \frac{\dot a}{a^2}\boldsymbol v_x + K \right] + \frac{1}{a^6} \boldsymbol \nabla_x A_x
\end{align*}
Combining all terms from the left-hand side of equation (\ref{momentumwithA}) gives
\begin{align*}
\frac{\partial( \rho \boldsymbol v)}{\partial t}\bigg|_r+\boldsymbol \nabla (\rho \boldsymbol v \boldsymbol v + A)&=\rho \frac{\partial \boldsymbol v}{\partial t}\bigg|_r+\boldsymbol v \frac{\partial \rho}{\partial t}\bigg|_r+\boldsymbol \nabla (\rho \boldsymbol v\boldsymbol v + A) \\
& = \frac{1}{a^3} \rho_x \left[ \frac{1}{a^3} \frac{\partial \boldsymbol v_x}{\partial t_x} + \frac{1}{a^3} (\boldsymbol v_x \cdot \boldsymbol \nabla_x) \boldsymbol v_x + \ddot a \boldsymbol x \right ] + \frac{1}{a^6} \boldsymbol \nabla_x A_x
\end{align*}
When comparing this to the right-hand side of equation (\ref{momentumwithA}),
\begin{align*}
-\rho \boldsymbol \nabla \phi &= \frac{-1}{a^4}\rho_x\boldsymbol \nabla_x \left( \frac{\phi_x}{a^2} - \frac{1}{2}a \ddot a x^2 \right) \\
&= \frac{-1}{a^6}\rho_x(\boldsymbol \nabla_x \phi_x) + \frac{1}{a^3}\rho_x \boldsymbol \nabla_x \left( \frac{1}{2} \ddot a  x^2 \right)\\
&= \frac{-1}{a^6}\rho_x(\boldsymbol \nabla_x \phi_x) + \frac{1}{a^3}\rho_x (\ddot a \boldsymbol x) \;\;\;,
\end{align*}
we notice that the $(\ddot a \boldsymbol x)$ term cancels, leaving
\begin{align}
\notag
\rho_x \frac{\partial \boldsymbol v_x}{\partial t_x} + &\rho_x ( \boldsymbol v_x \cdot \boldsymbol \nabla_x) \boldsymbol v_x + \boldsymbol \nabla_x A_x = -\rho_x \boldsymbol \nabla_x \phi_x \\
\notag
\Longleftrightarrow\;\;\;\;\;  \frac{\partial (\rho_x \boldsymbol v_x)}{\partial t_x} &+ \boldsymbol \nabla_x ( \rho_x \boldsymbol v_x  \boldsymbol v_x + A_x) = -\rho_x \boldsymbol \nabla_x \phi_x \\
\Longleftrightarrow\;\;\;\;\;  \frac{\partial (\rho_x \boldsymbol v_x)}{\partial t_x}& + \boldsymbol \nabla_x \left[ \rho_x \boldsymbol v_x  \boldsymbol v_x + \left(p_x+\frac{B_x^2}{2\mu}\right) I-\frac{1}{\mu} \boldsymbol B_x \boldsymbol B_x\right] = -\rho_x \boldsymbol \nabla_x \phi_x\;\;\;.
\end{align}


\subsection{Induction equation}
Before transforming the total energy density equation in the next subsection, the supercomoving induction equation has to be derived as it will be needed for it. We start from the equation
\begin{align*}
\frac{\partial \boldsymbol B}{\partial t}\bigg|_r + \boldsymbol \nabla \times (- \boldsymbol v \times \boldsymbol B) = 0
\end{align*}
subject to the condition
\begin{align*}
\boldsymbol \nabla \cdot \boldsymbol B = 0\;\;\;.
\end{align*}
The divergence-free condition turns out to be useful as it causes several terms to vanish. First we substitute the temporal and spatial derivatives:
\begin{align*}
&\frac{\partial \boldsymbol B}{\partial t}\bigg|_r + \boldsymbol \nabla \times (- \boldsymbol v \times \boldsymbol B) = 0 \\
\Longleftrightarrow\;\;\;\;\;&\frac{\partial \boldsymbol B}{\partial t}\bigg|_x - \frac{\dot a}{a} \boldsymbol x (\boldsymbol \nabla_x \cdot \boldsymbol B) +  \frac{1}{a}\boldsymbol \nabla_x \times \left[- \left( \frac{1}{a} \boldsymbol v_x + \dot a \boldsymbol x \right) \times \boldsymbol B\right] =0\\ 
\Longleftrightarrow\;\;\;\;\;&\frac{1}{a^2}\frac{\partial \boldsymbol B}{\partial t_x}  -  \frac{1}{a^2} \boldsymbol \nabla_x \times (\boldsymbol v \times \boldsymbol B) - \frac{\dot a }{a} \boldsymbol \nabla_x \times (\boldsymbol x \times \boldsymbol B) =0
\end{align*}
Again, the divergence-free condition allows us to simplify:
\begin{align*}
\boldsymbol \nabla_x \times (\boldsymbol x \times \boldsymbol B) = (\boldsymbol B \cdot \boldsymbol \nabla) \boldsymbol  x - \boldsymbol B (\boldsymbol \nabla \cdot\boldsymbol x) = \boldsymbol B - 3 \boldsymbol  B = -2 \boldsymbol  B
\end{align*}
and therefore
\begin{align*}
&\frac{1}{a^2}\frac{\partial \boldsymbol B}{\partial t_x}  -  \frac{1}{a^2} \boldsymbol \nabla_x \times (\boldsymbol v \times \boldsymbol B) + 2 \frac{\dot a}{a} \boldsymbol B=0 \\
\Longleftrightarrow\;\;\;\;\;&\frac{\partial \boldsymbol B}{\partial t_x}  -   \boldsymbol \nabla_x \times (\boldsymbol v \times \boldsymbol B) + 2 \dot a a \boldsymbol B=0
\end{align*}
Now we substitute the supercomoving magnetic field $\boldsymbol B_x = a^{5/2} \boldsymbol B$:
\begin{align*}
&a^{-5/2}\frac{\partial \boldsymbol B_x}{\partial t_x} + \boldsymbol B_x \frac{\partial}{\partial t_x}\left(a^{-5/2}\right) - a^{-5/2} \boldsymbol \nabla_x \times (\boldsymbol v_x \times \boldsymbol B_x) +2 \dot a a^{-3/2} \boldsymbol B_x = 0 \\
\Longleftrightarrow\;\;\;\;\;& a^{-5/2}\frac{\partial \boldsymbol B_x}{\partial t_x} - a^{-5/2} \boldsymbol \nabla_x \times (\boldsymbol v_x \times \boldsymbol B_x) - \frac{1}{2}\dot a a^{-3/2} \boldsymbol B_x = 0
\end{align*}
We define the supercomoving Hubble constant
\begin{align*}
\mathcal{H}:= \frac{1}{a} \frac{\textrm{d}a}{\textrm{d}t_x} = \dot a a
\end{align*}
With this notation, we have:
\begin{align}
\frac{\partial \boldsymbol B_x}{\partial t_x} + \boldsymbol \nabla_x \times (-\boldsymbol v_x \times \boldsymbol B_x) =\frac{1}{2} \mathcal{H}\boldsymbol B_x 
\end{align}

We defined the frame of reference such that it is comoving with the magnetic energy density, and not with the magnetic field strength. This is the reason why a magnetic Hubble drag term must appear at the right-hand side of the supercomoving induction equation and it is \emph{not} formally identical to the non-comoving induction equation. However, the Hubble term only ensures that the magnetic field scales properly with $a$; it does not have any physical meaning.


\subsection{Total energy density}
Instead of directly transforming the total energy equation (\ref{proper_f}), we derive the supercomoving energy conservation law by putting together all the quantities we have so far. The easiest way is to first derive the energy conservation for the hydrodynamic case and then add the magnetic energy density and flux to the result.

In the hydrodynamic case, $\rho E = \frac{1}{2}\rho v^2 + \rho\varepsilon$. One immediately notices from the definitions of the supercomoving variables that the supercomoving total energy is
\begin{align*}
\rho_x E_x = \frac{1}{2}\rho_x v_x^2 + \rho_x\varepsilon_x
\end{align*}
We start by calculating the temporal change of the kinetic energy  $\frac{1}{2}\rho_x v_x^2$ with the help of the already derived equations:
\begin{align*}
\frac{\partial}{\partial t_x}\left(\frac{1}{2}\rho_x v_x^2\right)&= \rho \boldsymbol v_x \frac{\partial \boldsymbol v_x}{\partial t_x} + \frac{1}{2}v_x^2\frac{\partial \rho_x}{\partial t_x} \\
& = \rho_x \boldsymbol v_x \left[ (\boldsymbol v_x \cdot \boldsymbol \nabla_x) \boldsymbol v_x - \boldsymbol \nabla_x \phi_x - \frac{1}{\rho_x} \boldsymbol \nabla_x p_x\right] - \frac{1}{2}v_x^2 \boldsymbol \nabla_x (\rho_x \boldsymbol v_x)  \displaybreak[0]\\
& = \rho_x \boldsymbol v_x \left[ -(\boldsymbol v_x \cdot \boldsymbol \nabla_x) \boldsymbol v_x \right] - \frac{1}{2}v_x^2\boldsymbol \nabla_x\cdot(\rho_x \boldsymbol v_x) - \ \boldsymbol v_x \cdot \boldsymbol \nabla_x p_x - \rho_x \boldsymbol v_x \cdot (\boldsymbol \nabla_x \phi_x) \displaybreak[0]\\
& = -\boldsymbol \nabla_x \cdot \left (\boldsymbol v_x \frac{1}{2}\rho_x v_x^2\right) - \boldsymbol v_x \cdot \boldsymbol \nabla_x p_x - \rho_x \boldsymbol v_x \cdot (\boldsymbol \nabla_x \phi_x)
\end{align*}
With $\boldsymbol v_x \cdot \boldsymbol p_x = \boldsymbol \nabla_x \cdot (\rho_x \boldsymbol v_x) - p_x \boldsymbol \nabla_x \cdot \boldsymbol v_x$ , we can write:
\begin{align}
\label{supercomovingkineticenergy}
\frac{\partial}{\partial t_x}\left(\frac{1}{2}\rho_x v_x^2\right) + \boldsymbol \nabla_x \cdot \left[ \left(\frac{1}{2}\rho_x v_x^2 + p_x\right) \boldsymbol v_x \right] = p_x \boldsymbol \nabla_x \cdot \boldsymbol v_x - \rho_x \boldsymbol v_x \cdot (\boldsymbol \nabla_x \phi_x)
\end{align}
Next, we need an equation for the thermal energy $\varepsilon_x$. In proper coordinates, such an equation exists (e.g. \citealt{Bryan95}). In the case of a monoatomic ideal gas ($\gamma=5/3$), which will be assumed from here on, it reads:
\begin{align}
\frac{\partial \varepsilon}{\partial t} + \boldsymbol v \cdot \boldsymbol \nabla \varepsilon = -\frac{1}{\rho}p \boldsymbol \nabla \cdot \boldsymbol v
\end{align}
By plugging in the definitions of the supercomoving variables it easily proves that the same equation holds in supercomoving coordinates:
\begin{align}
\frac{\partial \varepsilon_x}{\partial t_x} + \boldsymbol v_x \cdot \boldsymbol \nabla_x \varepsilon_x = -\frac{1}{\rho_x}p_x \boldsymbol \nabla_x \cdot \boldsymbol v_x
\end{align}
We can rewrite that as
\begin{align*}
p \boldsymbol \nabla_x \cdot \boldsymbol v_x &= - \rho_x \frac{\partial \varepsilon_x}{\partial t_x} - \rho \boldsymbol v_x \cdot \boldsymbol \nabla_x \varepsilon_x \\
&= -\frac{\partial (\rho_x \varepsilon_x)}{\partial t_x} - \boldsymbol \nabla_x [ \boldsymbol v_x (\rho_x \varepsilon_x) ]
\end{align*}
and plug it into equation (\ref{supercomovingkineticenergy}), yielding
\begin{align}
\label{supercomovinghydroenergy}
\frac{\partial }{\partial t_x}\left( \rho_x E_x\right) + \boldsymbol \nabla_x \cdot \left[ \left(\rho_x E_x+ p_x\right) \boldsymbol v_x \right] =  - \rho_x \boldsymbol v_x \cdot (\boldsymbol \nabla_x \phi_x)
\end{align}
This is the supercomoving total energy equation for the hydrodynamic case, again formally equivalent to the corresponding non-comoving equation.

Now we can consider the magnetic energy $B_x^2/2\mu$. Its temporal derivative is easily obtained from the supercomoving induction equation:
\begin{align*}
\frac{\partial}{\partial t_x} \left( \frac{B_x^2}{2\mu}\right)& = \frac{1}{\mu} \boldsymbol B_x \cdot \frac{\partial \boldsymbol B_x}{\partial t_x} \\
& =\frac{1}{\mu}\boldsymbol B_x \cdot \left [ \boldsymbol \nabla_x \times (\boldsymbol v_x \times \boldsymbol B_x ) + \frac{1}{2} \mathcal H \boldsymbol B_x \right ]\displaybreak[0] \\
& =\boldsymbol B_x \cdot \left [  \frac{1}{\mu} \boldsymbol \nabla_x \times (\boldsymbol v_x \times \boldsymbol B_x ) \right] + \mathcal H \frac{B_x^2}{2 \mu}\displaybreak[0] \\
& = \boldsymbol \nabla_x \cdot \left [  \frac{1}{\mu} (\boldsymbol v_x \times \boldsymbol B_x) \times \boldsymbol B_x \right] + \mathcal H \frac{B_x^2}{2 \mu} \\
& =\boldsymbol \nabla_x \cdot \left [ \frac{1}{\mu}  \boldsymbol B_x (\boldsymbol  v_x \cdot \boldsymbol B_x) - \boldsymbol v_x \frac{B_x^2}{2\mu} \right] + \mathcal H \frac{B_x^2}{2 \mu}
\end{align*}
Adding this equation to (\ref{supercomovinghydroenergy}), we get the supercomoving total energy equation for the full MHD case:
\begin{align}
\frac{\partial }{\partial t_x}\left( \rho_x E_x\right) + \boldsymbol \nabla_x \cdot \left[ \left(\rho_x E_x+ p_x\right) \boldsymbol v_x -  \frac{1}{\mu}  \boldsymbol B_x (\boldsymbol  v_x \cdot \boldsymbol B_x) \right] =  - \rho_x \boldsymbol v_x \cdot (\boldsymbol \nabla_x \phi_x)+ \mathcal H \frac{B_x^2}{2 \mu}
\end{align}
where now
\begin{align*}
\rho_x E_x = \frac{1}{2}\rho_x v_x^2 + \rho_x\varepsilon_x + \frac{B_x^2}{2\mu}\;\;\;.
\end{align*}

\subsection{Modified entropy}

This additional equation is needed to use the ``S system'' in the dual energy formalism. Transforming the first term:

\begin{align*}
\frac{\partial S}{\partial t}\bigg|_r \rho&=\frac{\partial S}{\partial t}\bigg|_x \rho-\frac{\dot a}{a}\boldsymbol x\cdot \boldsymbol \nabla_x S\\
&=\frac{1}{a^2}\frac{\partial}{\partial t_x} \left( a^{3\gamma -8} S_x\right)-\frac{\dot a}{a}\boldsymbol x\cdot\boldsymbol \nabla_x \left( a^{3\gamma - 8}S_x\right) \displaybreak[0]\\
&=a^{3\gamma-10} \left[\frac{\partial S_x}{\partial t_x} + \mathcal H (3\gamma-8) S_x - \mathcal H \boldsymbol x \cdot \boldsymbol \nabla_x S_x \right]
\end{align*}
and the second term:
\begin{align*}
\boldsymbol \nabla \cdot (S \boldsymbol v) &= \frac{1}{a} \boldsymbol \nabla_x \cdot \left[ a^{3\gamma -8} S_x \left( \frac{1}{a} \boldsymbol v_x + \dot a \boldsymbol x \right) \right ] \\
&= a^{3\gamma-10} \boldsymbol \nabla_x \cdot \left[ S_x \boldsymbol v_x + \mathcal H \boldsymbol x S_x \right] \\
&= a^{3\gamma-10} \left[ \boldsymbol \nabla_x \cdot \left( S_x \boldsymbol v_x \right) + \mathcal H \boldsymbol x \cdot \boldsymbol \nabla_x S_x + 3 \mathcal H S_x \right]
\end{align*}
Putting both together yields
\begin{align}
\notag
&\frac{\partial S_x}{\partial t_x} + \mathcal H(3\gamma-8) S_x + \boldsymbol \nabla_x \cdot (S_x \boldsymbol v_x) + 3\mathcal HS_x =0 \\
\Longleftrightarrow\;\;\;\;\; &\frac{\partial S_x}{\partial t_x} + \boldsymbol \nabla_x \cdot (S_x \boldsymbol v_x) = -\mathcal H (3\gamma-5)\;\;\;\;\;\;.
\end{align}

\end{onecolumn}
\end{document}